\documentclass{aastex631}

\usepackage{graphicx}	
\usepackage{amsmath}

\shorttitle{Low-frequency Radio Emission from 7 Nearby Large Galaxies}
\shortauthors{Manna, Roy, and Baug}

\begin{document}

\title{Radio Continuum Halos of 7 Nearby Large Galaxies using uGMRT}


\correspondingauthor{Souvik Manna}
\email{souvik.manna@bose.res.in, souvik@ncra.tifr.res.in}

\author{Souvik Manna}
\affiliation{S. N. Bose National Centre for Basic Sciences, \\ Block-JD, Sector-III, Salt Lake, Kolkata 700106, India}
\affiliation{National Center for Radio Astrophysics, TIFR, \\ Pune University Campus, Ganeshkhind, Pune 411007, India}

\author{Subhashis Roy}
\affiliation{National Center for Radio Astrophysics, TIFR, \\ Pune University Campus, Ganeshkhind, Pune 411007, India}

\author{Tapas Baug}
\affiliation{S. N. Bose National Centre for Basic Sciences, \\ Block-JD, Sector-III, Salt Lake, Kolkata 700106, India}


\begin{abstract}
We present the results of deep radio observations of 7 nearby large galaxies observed using the upgraded Giant Metrewave Radio Telescope (uGMRT) 0.3-0.5 GHz receivers with an angular resolution of $\sim$10 arcsec. The achieved sensitivities of these observations range from $\approx$15 to 50 $\mu$Jy/beam which is $\approx$3-4 factor lower than the previous observations at these frequencies. For 2 galaxies (NGC3344 and NGC3627) with moderate inclination angles, significant diffuse emissions are seen for the first time. Detected radio halos in the vertical direction are significantly larger in our 0.4 GHz maps than compared to the observations at $\sim$1.5 GHz for 4 nearly edge-on galaxies - NGC3623, NGC4096, NGC4594, and NGC4631. For these 4 galaxies, significantly larger halos are also detected along the galaxy disk. For NGC3623 and NGC4594, we could detect elongated radio disks which was not seen before.  
We also present new uGMRT images of NGC3344 and NGC3623 at 1.3 GHz and a new VLA image of NGC3627 at 1.5 GHz. We fitted an exponential function to the flux densities along different cross-cuts and found a significantly wider distribution at 0.4 GHz uGMRT images than compared to the high-frequency images at $\sim$1.5 GHz. Using maps at 0.144, 0.4, and $\sim$1.5 GHz, we made spectral index maps of the 7 sample galaxies and found steepening of the spectrum up to a value of $\sim$ -1.5 in the halo regions of the galaxies. 

  
\end{abstract}



\section{Introduction}
\label{introduction}

A connection between the extra-planar halo and the Interstellar Medium (ISM) of the underlying disk region has been reported in the past using observations in a wide range of frequency bands from X-ray to radio \citep[e.g.][]{Dahlem1995ApJ, Rossa2003A&A, Strong2004A&A, Tullmann2006A&A}. 
Radio-frequency observations also reveal that the propagation scale lengths of Cosmic Ray Electrons (CREs) in the vertical direction are larger than the scale length in the disk region \citep[e.g.][]{Dahlem2006A&A, Mulcahy2014A&A} providing evidence of the transport of CREs from the disk into the halo region. 
Besides, an X-shaped structure of the magnetic field has recently been observed in the halo region by stacking the polarized emission from a sample of 16 nearby edge-on spiral galaxies using L- and C-band VLA observations \citep[CHANG-ES Survey,][]{Krause2020A&A}.
Transport of cosmic rays in most of the CHANG-ES galaxies were found to be dominated by advection and the role of magnetic fields on such transport remains unknown \citep{Irwin2024}.
Magnetic fields and CREs are key components to study these disk-halo connections as they are believed to have a critical role in driving outflows \citep[e.g.][]{Ipavich1975ApJ, Everett2010ApJ}. Therefore, it is important to study the distribution of magnetic fields and CREs in different types of nearby galaxies to better understand if there exist
large differences between individual galaxies in the above context and also if these differences could constrain theoretical models of CRE propagation.


Low radio-frequency observations are excellent probes to characterize magnetic fields and CREs of nearby galaxies as non-thermal synchrotron emission, which traces both the components, dominates at low radio-frequencies due to its steep spectral index ($\approx$0.8). The contribution of the thermal free-free component to the total observed radio emission becomes small ($\sim$ 5\%) at frequencies below 1 GHz \citep{Basu2012MNRAS, RoyManna2021}. However, most of the studies on different types of nearby galaxies have been carried out at frequencies above 1 GHz. Although there exist observations of nearby galaxies below 1 GHz, they were mostly made with poor sensitivity (rms noise $\sim$1mJy/beam). Therefore, new observations at low-radio frequency ($<$1GHz) with high resolution and high sensitivity are essential to better understand the connection between the disk and the halo of nearby galaxies. 

We started the $``$Metrewavelength survey of Local Volume Large galaxies$"$ (MLVL), a GMRT survey of 46 nearby galaxies at $\approx$0.4~GHz, intending to characterize the distribution of magnetic fields and CREs in nearby galaxies at spatial resolutions below 1 kpc. As a pilot project, we observed 7 MLVL galaxies with the legacy GMRT 0.325~GHz receivers and investigated the dominant physical processes related to radio emission within the sample galaxies by studying (1) the radio-IR correlations, (2) the scale lengths of CRE propagation, (3) the correlations between magnetic fields, gas densities, and star-formation rates \citep{RoyManna2021,Manna2023ApJ} at the sub-kpc linear resolution. 
We further observed 7 nearby galaxies, mostly edge-on, from our MLVL sample using the uGMRT 0.3-0.5 GHz receivers, which is expected to provide significantly better sensitivity ($\approx$3 times) than the legacy GMRT.
 Among these galaxies, NGC3623, NGC3628, NGC4096, NGC4594, and NGC4631, are nearly edge-on with inclination angles greater than 75 degrees. The other 2 galaxies, NGC3344, and NGC3627 have inclination angles of 64 and 65 degrees respectively.

In this paper, we present the high- and low-resolution images of the 7 sample galaxies at 0.4 GHz observed as a part of our MLVL survey using the uGMRT. We also present uGMRT images of 2 galaxies at 1.3 GHz observed as a part of the MLVL survey. Besides, we present VLA wideband image of NGC3627 at 1.5 GHz which we analyzed using archival data. We also incorporated images of 4 galaxies at $\sim$1.5 GHz, from the two surveys titled $``$Continuum Halos in Nearby Galaxies (CHANG-ES)$"$ \citep{Irwin2012AJ144} and $``$LOFAR Two-metre Sky Survey (LoTSS)$"$ \citep{Shimwell2022A&A} using the VLA and the Low-Frequency Array (LOFAR) respectively, for which publicly available images from these surveys are available. We compare both images at 0.4 and $\sim$1.5 GHz by measuring the size of the radio halos and also by fitting an exponential function to the flux densities. We also made spectral index maps of all the sample galaxies and discussed the steepening of the spectrum in the halo regions. 

This paper is organized as follows. Details of the sample galaxies are discussed in Section \ref{sample}. We describe the uGMRT observations and the data analysis in Section \ref{data_analysis}. In Section \ref{results} and \ref{discussion}, we present the results and discussions of our study. Finally, we summarize our work in Section \ref{summary}.


\section{Sample}
\label{sample}
 The 46 galaxies of the MLVL survey are drawn from a complete volume-limited survey, the Local Volume Legacy (LVL) survey, containing 258 nearby galaxies within a distance of 11 Mpc \citep{Dale2009ApJ}.
We selected the galaxies with angular sizes larger than 6 arcmin and smaller than 17 arcmin. The former criterion was chosen to have a large number of beams across the galaxies whereas the latter one was to minimize the effect of missing flux at short-spacing for an interferometer like GMRT \citep{swarup91gmrt}. 
This sample selection criteria of 46 galaxies have also been discussed in detail in \cite{RoyManna2021}.
Existing multi-wavelength observations of the LVL sample at UV, H$\alpha$, optical, and multi-frequency IR bands make radio observations highly complementary to those observations.

In Table \ref{sample_details}, we present morphological types, distances, inclination, and position angles of these 7 sample galaxies. 
The distances and the position angles are taken from \cite{Dale2009ApJ}. Detailed discussions on individual galaxies are presented in Section \ref{Individual_Galaxy}.


\begin{table}
 \caption{Details of the 7 sample galaxies. Inclination angles (to the sky plane) of the galaxies are from 
 (1) \cite{Verdes2000A&A}, (2) \cite{Burbidge1961ApJ}, (3) \cite{RoyManna2021}, (4) \cite{Baes2016A&A}, (5) \cite{RoyManna2021}, (6) \cite{Wiegert2015AJ} and (7) \cite{Irwin2011MNRAS}.}
\scriptsize
\centering
 \begin{tabular}{||c c c c c||} 
 \hline
 
Name & Morphological & Distance & Inclination  & Position \\ 
     &   class       &   (Mpc)  &  angle             &  angle \\
     
 & & &  (deg)& (deg) \\
 
 \hline
 
NGC3344 & (R)SAB(r)bc & 6.64 & 64$^{1}$ & 330  \\ 
NGC3623       & SAB(rs)a & 8.95 & 75$^{2}$ & 352    \\
NGC3627       & SAB(s)b & 10.05 & 65$^{3}$ &  347   \\
NGC3628       & Sb      & 9.45  & 88$^{4}$ &  102 \\
NGC4096       & SAB(rs)c & 8.28 & 76$^{5}$ & 18   \\
NGC4594       & SA(s)a & 9.33 & $>$75$^{6}$ & 90   \\
NGC4631 & SB(s)d  & 8.05 & 86$^{7}$ & 80  \\
 
 \hline
 \end{tabular}
\label{sample_details}
\end{table}
\section{Observations and Data Analysis}
\label{data_analysis}
\subsection{Data at 0.4 GHz}
\label{band3_data}

Observations of the 7 sample galaxies were carried out using uGMRT \citep{Gupta2017CSci} band-3 receivers (0.3-0.5 GHz) from 18th January 2019 to 3rd March 2019 (Project: 35\_113).
A bandwidth of 200 MHz with an on-source time of 3.3 to 7.5 hours were used to observe each of these sources.
Details of the observations are listed in Table \ref{obs_details}. We used a total of 1024 spectral channels across the observing band with visibilities recorded with an integration time of $\sim$11 seconds. Flux calibrators (3C48, 3C147, or 3C286) were observed at least once near the middle of each observing session, for $\approx$5 minutes. The phase calibrators (which were also used for bandpass calibration) and the target source were observed one after the other forming a loop for a time of $\approx$3.5 and $\approx$25 minutes respectively. 
We typically observed 2 galaxies in each observing session of 8-12 hours in multiple snapshot modes to improve the uv-coverage.
All the observations were conducted at night to reduce Radio Frequency Interference (RFI).  

The data were analyzed using the Astronomical Image Processing System (AIPS) \citep{Wells1985daa}, and the Common Astronomy Software Applications (CASA) \citep{McMullin2007ASPC}. We performed the standard procedure for the initial flagging, gain calibration, and bandpass calibration of the data using the AIPS.
Additionally, we used AOFlagger \citep{Offringa012A&A} to flag the wideband data of the target source. 
The imaging and self-calibration of the wideband data were carried out following the standard approach in CASA. We excluded the Central Square Baselines ($\leq$1 km) during the self-calibration procedure as these baselines are likely to be more affected by RFI. 
After the self-calibration was done, we used the casa task UVSUB to subtract the clean components from the original self-calibrated uv-data and flagged the remaining RFI on the residual uv-data using AOFlagger. Subsequently, we made the final images using casa task TCLEAN with a robust parameter of 0.5.
We also made low-resolution images, at resolutions of $\sim$15-20 arcsecs, of the sample galaxies to achieve better sensitivity for the low-surface brightness emissions (diffuse emission). The above procedure was performed by applying an upper cutoff to the uvrange. 

\subsection{Data at $\sim$1.4GHz}
\label{band5_data}

We aimed to make spectral index maps for the sample galaxies and therefore, radio images at higher frequencies with resolutions and sensitivities comparable to our 0.4 GHz maps were required. Four of our sample galaxies, (NGC3628, NGC4096, NGC4594, and NGC4631) have VLA data at the L band in different array configurations \citep[CHANG-ES Survey,][]{Irwin2012AJ144}. We used the L-band C-array images which have comparable resolutions and sensitivity with our uGMRT maps. Besides, 2 of our galaxies, NGC4096 and NGC4631, have archival LOFAR data at 0.144 GHz from the LoTSS survey \citep{Shimwell2022A&A}. We used data at 1.5 and 0.144 GHz with our data at 0.4 GHz to make spectral index maps of these 4 galaxies. 

NGC3627 was observed using the VLA C-array configuration (Project code: 17A-396, observing date: 28 June 2017) with a bandwidth of 256~MHz and an on-source time of $\approx$22 minutes. We used the $``$VLA calibration pipeline$"$ and the $``$VLA imaging pipeline$"$ to analyze this data. 

For the other 2 galaxies (NGC3344 and NGC3623), deep high-frequency radio observations were not available. 
We, therefore, exploited the band-5 receivers of the uGMRT to propose new observations of these 2 galaxies at 1.2-1.4 GHz. The uGMRT observations were made in observing cycle 39 from February to March 2021 (Project: 39\_098). Details of the band-5 observations such as bandwidth, on-source time, and the number of working antennas are provided in Table \ref{obs_details_band5}. Similar to the band-3 data, the standard procedures for the initial flagging, gain calibration, and bandpass calibration of the data were performed using AIPS.
We also used AOFlagger to flag the wideband data of the target source. We performed the imaging and self-calibration of the target source following the standard process using AIPS. These images were made with resolutions comparable to the previously made low-resolution images at 0.4 GHz. Final imaging was done was done with $``$wsclean$"$ \citep{Offringa2017MNRAS} using W-gridding.

Flux density values greater than 4 times the rms noise of the maps at 0.144, 0.4, and 1.4 GHz were used to make the spectral index maps of all the sample galaxies. 

\begin{table}
 \caption{Details of the uGMRT observations at 0.3-0.5 GHz.}
\scriptsize
\centering
 \begin{tabular}{||c c c c c c c||} 
 \hline

Name & Right ascension & Declination & Observing Date & On-source time & bandwidth  & No. of    \\ 
     & (RA)             & (DEC)                       &  & (hour)      &  (MHz)     & Working antennas   \\
 
 \hline
 
NGC3344 & 10h43m31.15s & +24d55m20.0s & 2019 January 18 & 3.33 & 200 & 30  \\ 
NGC3623  & 11h18m55.916s & +13d05m32.51s & 2019 February 11 & 7.00 & 200 & 30     \\
NGC3627  & 11h20m14.964s & +12d59m29.54s & 2019 February 11 & 7.00 & 200 & 30    \\
NGC3628  & 11h20m16.970s & +13d35m22.86s & 2019 February 11 & 7.00 & 200 & 30    \\
NGC4096 & 12h06m01.130s & +47d28m42.40s & 2019 March 02    & 4.17 & 200 & 30  \\
NGC4594 & 12h39m59.4318s & -11d37m22.996s & 2019 March 03    & 3.75 & 200 & 30    \\
NGC4631 & 12h42m08.01s & +32d32m29.4s &  2019 February 22 & 7.50 & 200 & 30   \\
 
\hline
\end{tabular}
\label{obs_details}
\end{table}

\begin{table}
 \caption{Details of the uGMRT observations at 1.2-1.4 GHz.}
\scriptsize
\centering
 \begin{tabular}{||c c c c c||} 
 \hline

Name & Observing Date & On-source time & bandwidth  & No. of    \\ 
     &                & (hour)         &  (MHz)     & Working antennas   \\
 
 \hline
 
NGC3344  & 2021 January 15 & $\sim$3 & 200 & 30    \\
NGC3623  & 2021 January 15 & $\sim$3 & 200 & 30  \\ 

\hline
\end{tabular}
\label{obs_details_band5}
\end{table}

\begin{table}
 \caption{Details of the high- and low- resolution images. The rms noise with superscript 1 is the mean rms noise at regions far away from the field centre. Similarly, the rms noise with superscript 2 is the mean rms noise at regions surrounding the galaxy.}
\scriptsize
\centering
 \begin{tabular}{||c c c c c c c c c c c||} 
 \hline

 &  & & High-resolution  & & & & & Low-resolution & &\\
 \hline
Name &  uv   & Angular    & rms$^{1}$ & rms$^{2}$  & Total &  uv & Angular  & rms$^{1}$ & rms$^{2}$ & Total \\ 
     & range & resolution & noise     & noise  & flux &  range & resolution & noise & noise & flux \\
     
     & (k$\lambda$) & $(\textrm{arcsec}^{2})$  & ($\mu$Jy/beam) & ($\mu$Jy/beam) & (mJy) &  (k$\lambda$) & $(\textrm{arcsec}^{2})$ & ($\mu$Jy/beam) & ($\mu$Jy/beam) & (mJy) \\
 
 \hline
 
NGC3344   & 0.06-30.0 & 10$\times$8  & 22 & 50 & 213$\pm$6 & 0.06-5.0 & 23$\times$22 & 50 & 100 & 265$\pm$14 \\ 
NGC3623   & 0.1-26.0 & 11$\times$10 & 35 & 120 & 76$\pm$18 & 0.1-10.0 & 17$\times$15 & 50  & 175 &  78$\pm$20  \\
NGC3627   & 0.1-26.0 & 11$\times$10 & 35 & 120 & 1280$\pm$110 & 0.1-10.0 & 17$\times$15 & 50 & 180 &  1370$\pm$140 \\
NGC3628   & 0.1-26.0 & 11$\times$10 & 35 & 130 & 1170$\pm$200 & 0.1-10.0 & 17$\times$15 &  50   & 200 & 1200$\pm$250 \\
NGC4096   & 0.11-25.0 & 10$\times$9 & 15 & 35 & 144$\pm$10 & 0.11-10.0 & 15$\times$14 & 22   & 50 &  137$\pm$12 \\
NGC4594   & 0.1-26.0 & 10$\times$9  & 23 & 55 & 126$\pm$35 & 0.1-10.0 & 18$\times$15 &  38 & 70 & 118$\pm$37 \\
NGC4631   & 0.06-26.0 & 10$\times$9  & 20 & 70 & 2760$\pm$70 & 0.06-10.0 & 16$\times$14 &  30   & 100 & 2690$\pm$140 \\

 \hline
 \end{tabular}
\label{uGMRT_map_details}
\end{table}

 \section{Results}
 \label{results}
\begin{figure}
 \centering
\includegraphics[trim={0 0 0 0},clip,scale=0.3,width=0.45\linewidth]{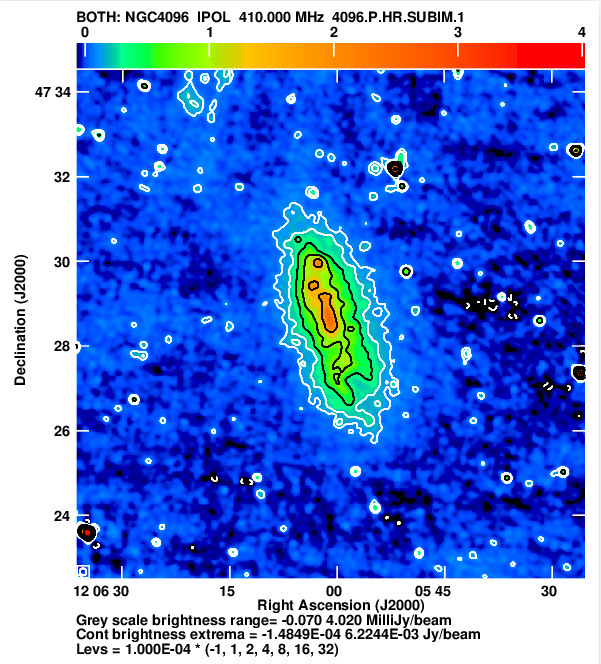}
\includegraphics[trim={0 0 0 0},clip,scale=0.3,width=0.45\linewidth]{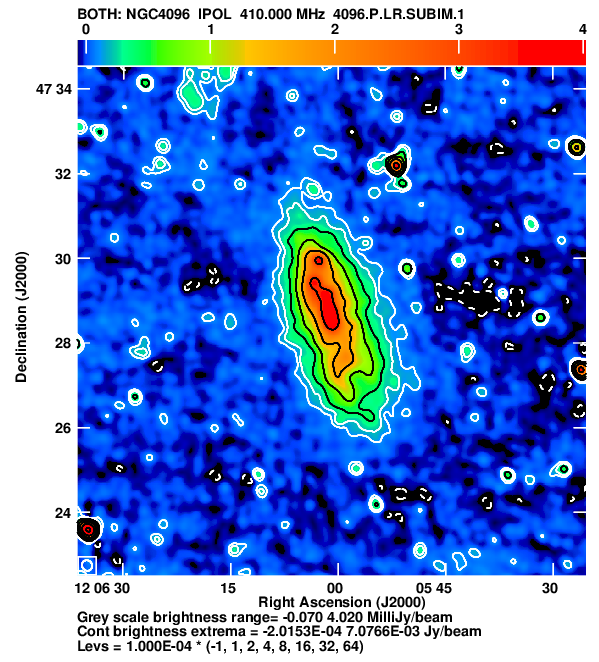}
 \includegraphics[trim={0 0 0 0},clip,scale=0.3,width=0.4\linewidth]{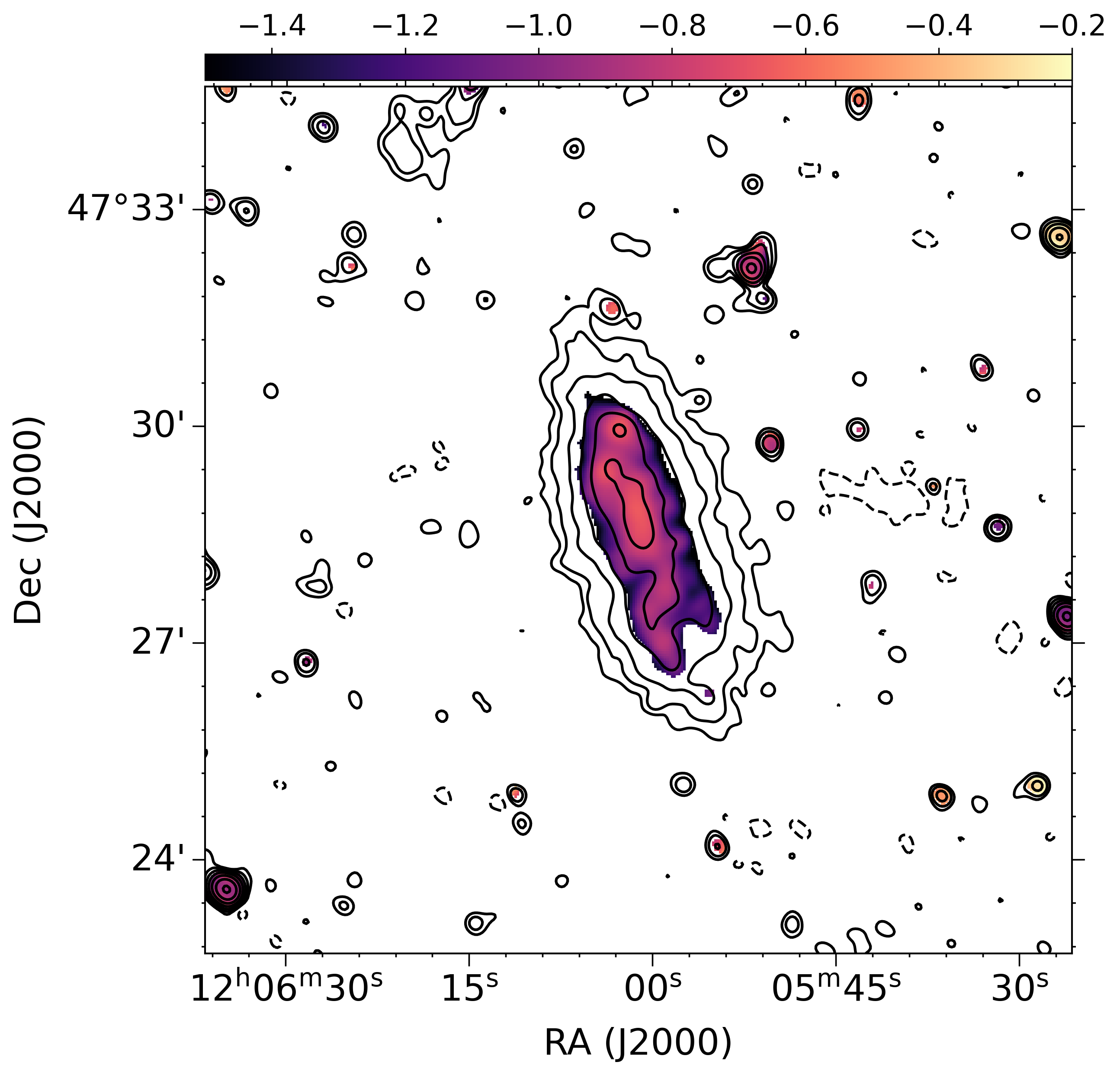}
\includegraphics[trim={0 0 0 0},clip,scale=0.3,width=0.4\linewidth]{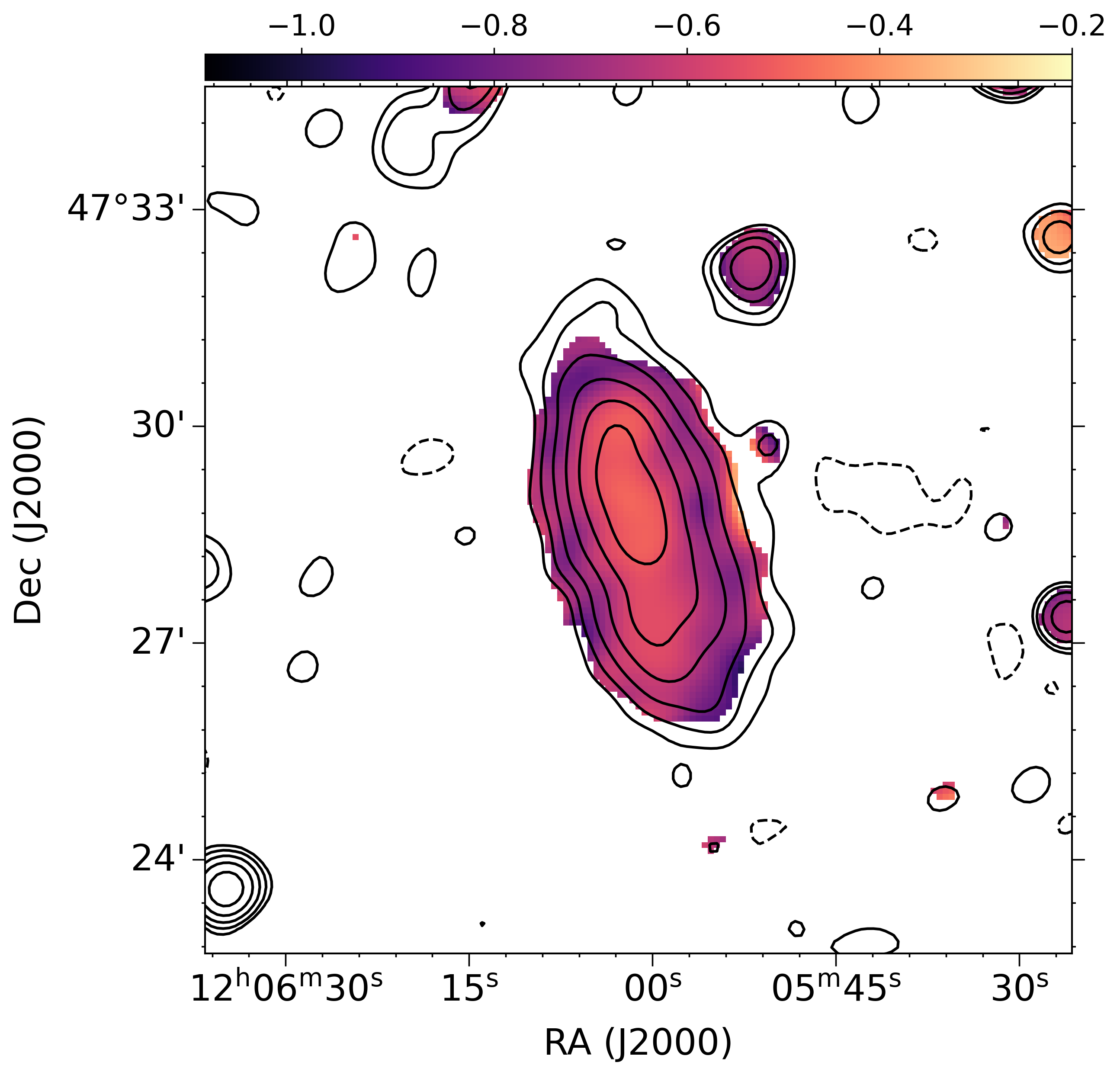}
\caption{Top two panels show the high- and low-resolution image of NGC4096 at 0.4 GHz in contours and color scales. The contour levels are indicated below the figure. The resolution of the high- and low-resolution maps are provided in Table \ref{uGMRT_map_details}. The bottom left panel shows the spectral index map using 0.4 GHz uGMRT and 1.5 GHz VLA image \citep{Irwin2012AJ144}, in colors overlaid on the contours of the 0.4 GHz uGMRT image. The contour levels of the bottom left panel are plotted at 1.0$\times$10$^{-4}$ (-1, 1, 2, 4, 8, 16, 32, 64, 128, 256) Jy/beam. The bottom right panel shows the spectral index map using 0.144 GHz \citep{Shimwell2022A&A} and 1.5 GHz maps, in colors overlaid on the contours of the 0.4 GHz uGMRT image. The contour levels of the bottom right panel are plotted at 3.0$\times$10$^{-4}$ (-1, 1, 2, 4, 8, 16, 32, 64, 128, 256) Jy/beam. The spectral index maps have resolutions of 15$\times$14 arcsec$^{2}$ and 33$\times$32 arcsec$^{2}$ respectively. The spectral index maps have mean errors of 0.03 and 0.08 respectively.}
\label{gwb_contours_4096}
\end{figure}

\begin{figure}
 \centering
\includegraphics[trim={0 0 0 0},clip,scale=0.3,width=0.45\linewidth]{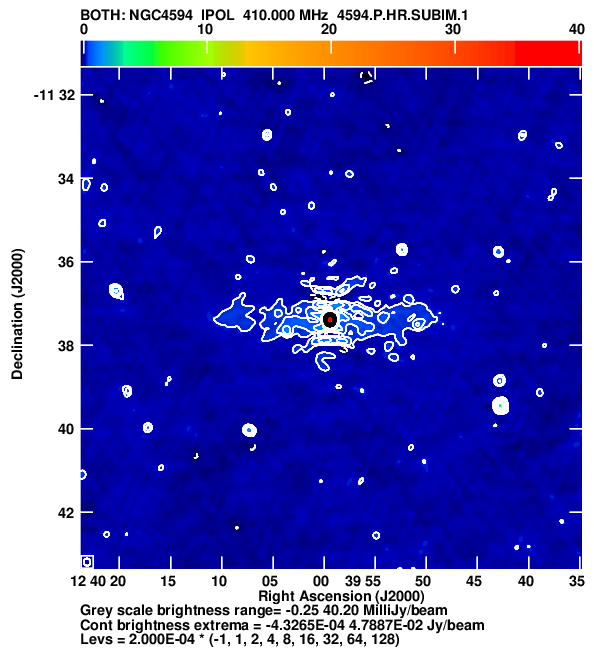}
\includegraphics[trim={0 0 0 0},clip,scale=0.3,width=0.45\linewidth]{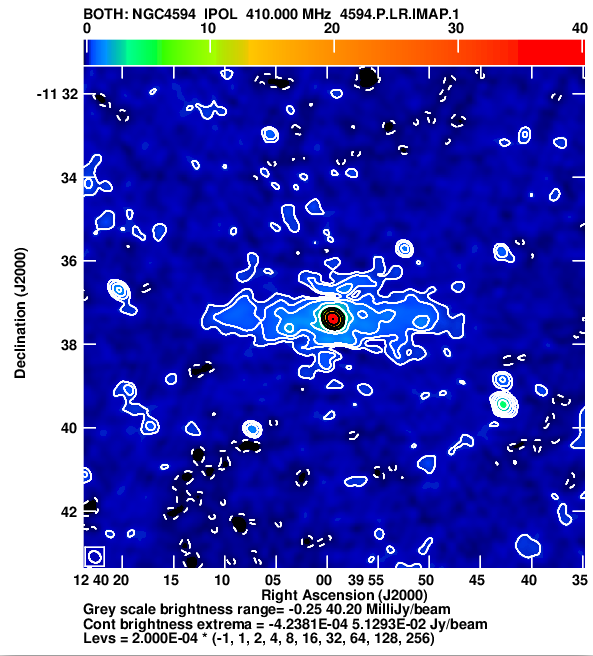}
\includegraphics[trim={0 0 0 0},clip,scale=0.3,width=0.4\linewidth]{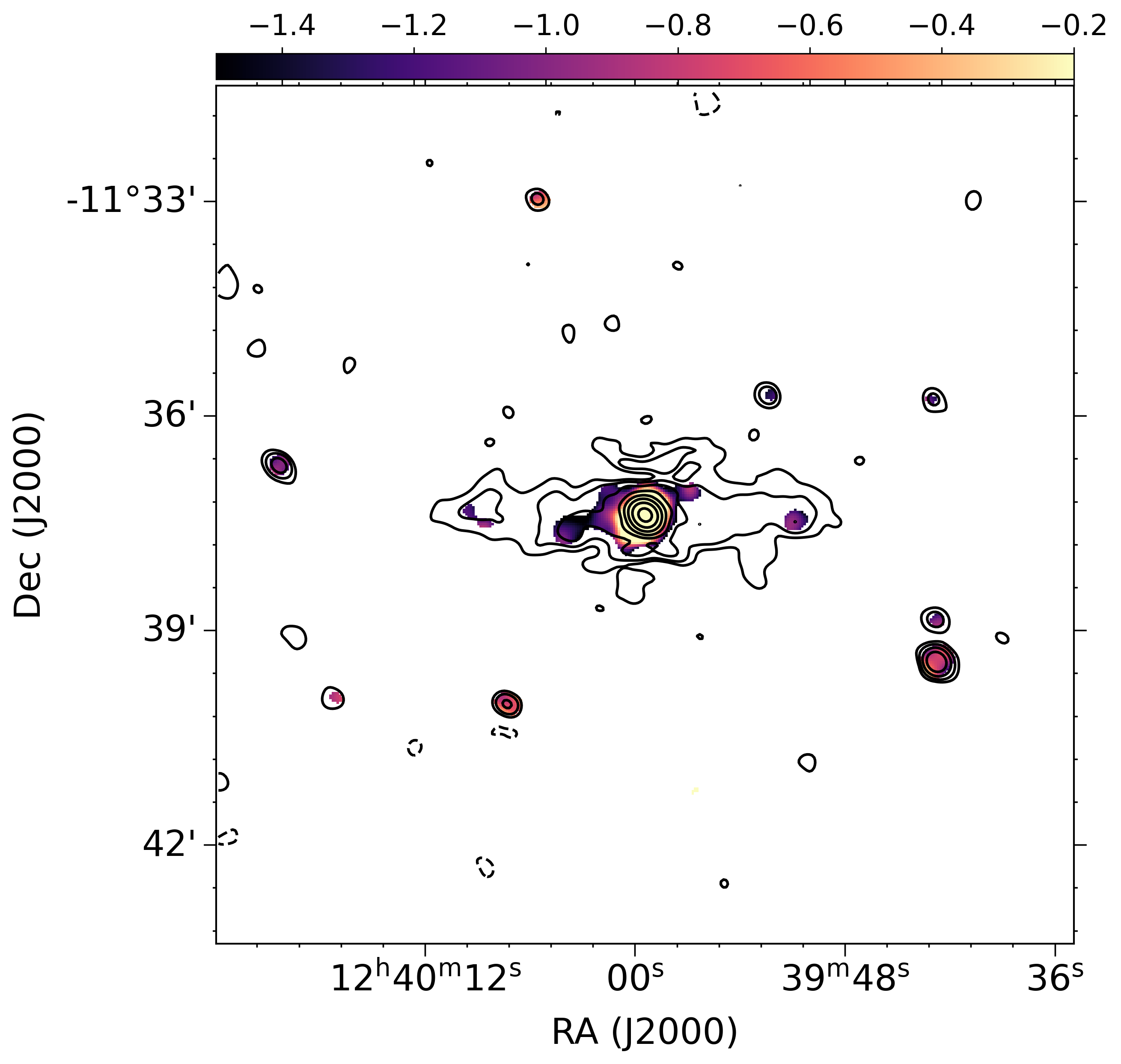}
\caption{Top two panels show the high- and low-resolution image of NGC4594 at 0.4 GHz in contours and color scales. The contour levels are indicated below the figure. The resolution of the high- and low-resolution maps are provided in Table \ref{uGMRT_map_details}. The bottom panel shows the spectral index map between 0.4 GHz and 1.5 GHz \citep{Irwin2012AJ144} in colors overlaid on the contours of the 0.4 GHz uGMRT image. The contour levels of the bottom panel are plotted at 3.0$\times$10$^{-4}$ (-1, 1, 2, 4, 8, 16, 32, 64, 128, 256) Jy/beam. The spectral index map has a resolution of 18$\times$15 arcsec$^{2}$ and a mean error of 0.09.}
\label{gwb_contours_4594}
 \end{figure}

We present high-resolution ($\sim$10 arcsec) contour images of the 7 sample galaxies at 0.3-0.5 GHz in Figure \ref{gwb_contours_3344}, \ref{gwb_contours_3623}, \ref{gwb_contours_3627}, \ref{gwb_contours_3628}, \ref{gwb_contours_4096}, \ref{gwb_contours_4594} and \ref{gwb_contours_4631}. Flux densities are shown in both color scales and contours. Contour levels are provided below each panel of the figure.
In Table \ref{uGMRT_map_details}, we present (1) the uv-range used for the high-resolution imaging, (2) the achieved angular resolution, (3) the mean rms noise of the observing field, (4) the mean rms noise in the surrounding region of the galaxy, and, (5) the galaxy-integrated flux density. The errors on the flux densities quoted in this paper only provide the uncertainties from image rms. To find the overall errors, one needs to add in quadrature another $\sim$10\% error in flux densities.

We made low-resolution images of these galaxies, using a lower uv-range (Table \ref{uGMRT_map_details}), to detect more diffuse emissions than the high-resolution ones. Contour maps of the low-resolution images of the sample galaxies are presented in Figures \ref{gwb_contours_3344}, \ref{gwb_contours_3623}, \ref{gwb_contours_3627}, \ref{gwb_contours_3628}, \ref{gwb_contours_4096}, \ref{gwb_contours_4594} and \ref{gwb_contours_4631}. Note that both the contour maps at high- and low- resolution are made with the same regions, same contour levels, and the same pixranges. Details of these images are provided in Table \ref{uGMRT_map_details}.

We also provide uGMRT images of 2 galaxies (NGC3344 and NGC3623) at 1.3 GHz and VLA wideband image of NGC3627 at 1.5 GHz, which are made to study the spatially-resolved spectrum of the sample galaxies. Contour maps of these 3 galaxies are shown in Figures \ref{gwb_contours_3344}, \ref{gwb_contours_3623}, and \ref{gwb_contours_3627}. Similar to the contour images at 0.4 GHz, we present flux densities in both colors and contours. These images are made at resolutions comparable to the low-resolution images at 0.4 GHz. 
We achieved mean rms noises of 100, 100, and 150 $\mu$Jy/beam in the surrounding region of NGC3344, NGC3623, and NGC3627 respectively. The achieved resolutions for these 3 galaxies are 22$\times$22, 15$\times$15, 28$\times$13 arcsec$^{2}$ respectively.

We present spectral index maps of the 7 galaxies in Figures \ref{gwb_contours_3344}, \ref{gwb_contours_3623}, \ref{gwb_contours_3627}, \ref{gwb_contours_3628}, \ref{gwb_contours_4096}, \ref{gwb_contours_4594} and \ref{gwb_contours_4631}. Flux-density contours of 0.4 GHz observations are overlaid on the spectral index maps. 
We discuss results from the individual galaxies in the following sub-section.

\subsection{Individual Galaxy}
\label{Individual_Galaxy}

\begin{figure}
 \centering
\includegraphics[trim={0 0 0 0},clip,scale=0.3,width=0.45\linewidth]{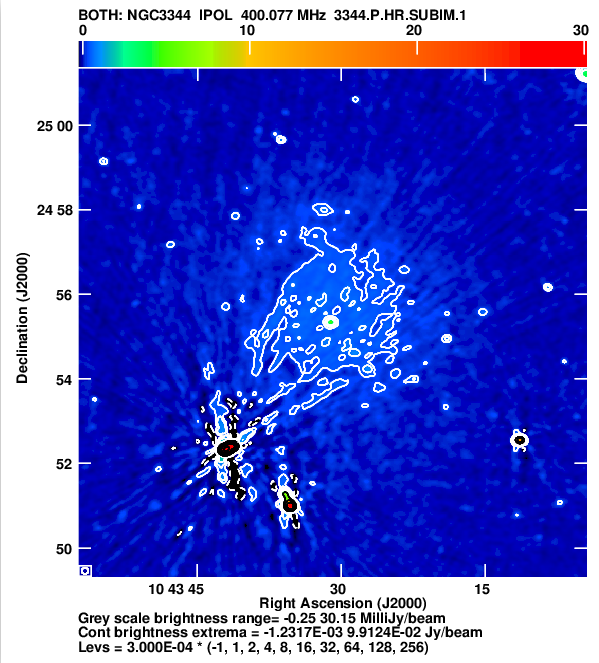}
\includegraphics[trim={0 0 0 0},clip,scale=0.3,width=0.45\linewidth]{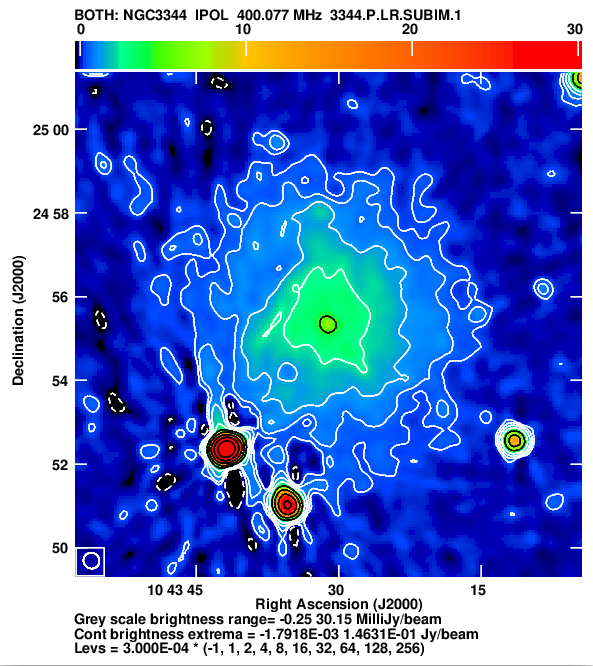}
\includegraphics[trim={0 0 0 0},clip,scale=0.3,width=0.45\linewidth]{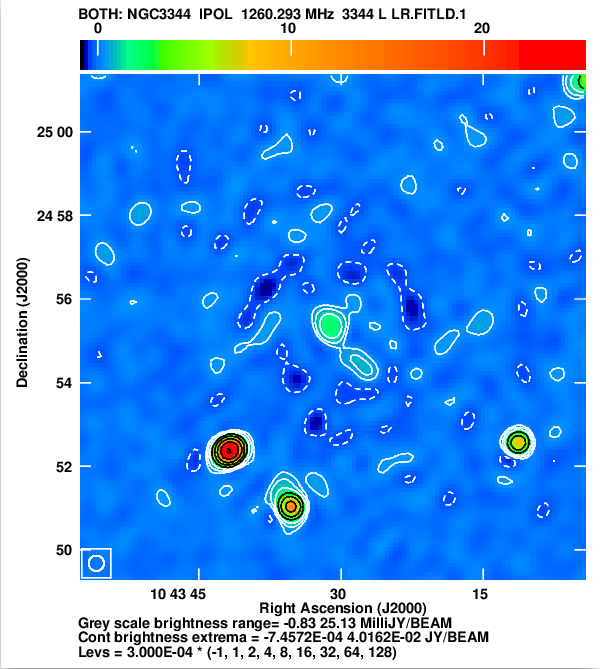}
 \includegraphics[trim={0 0 0 0},clip,scale=0.3,width=0.45\linewidth]{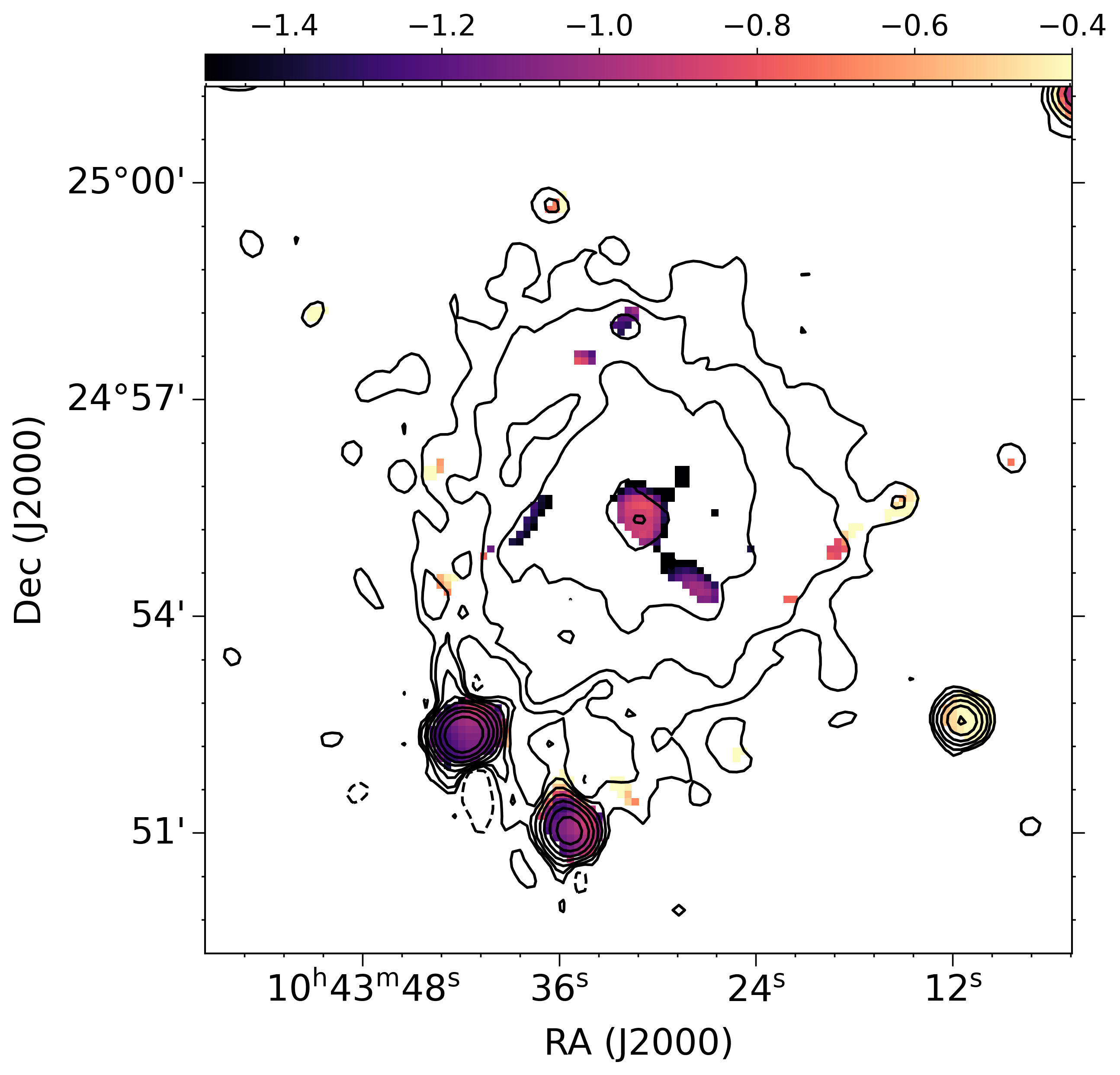}
\caption{Top two panels show the high- and low-resolution image of NGC3344 at 0.4 GHz in contours and color scales. The contour levels are indicated below the figure. The resolution of the high- and low-resolution maps are provided in Table \ref{uGMRT_map_details}. The bottom left panel shows the uGMRT contour image at 1.3 GHz where the flux densities are shown in colors and contours. The bottom right panel shows the spectral index map between 0.4 GHz and 1.3 GHz in colors overlaid on the contours of the 0.4 GHz uGMRT image. The contour levels of the bottom right panel are plotted at 4.0$\times$10$^{-4}$ (-1, 1, 2, 4, 8, 16, 32, 64, 128, 256) Jy/beam. The spectral index map has a resolution of 23$\times$22 arcsec$^{2}$ and a mean error of 0.08.}
\label{gwb_contours_3344}
 \end{figure}

{\bf NGC3344} is an isolated SABbc-type galaxy composed of a small bar, an inner ring, and an outer ring \citep{Verdes2000A&A}. NGC3344 has an inclination angle of 64 degrees \citep{Verdes2000A&A}. 
\cite{Condon1987} observed this galaxy using VLA at 1.49 GHz and estimated the galaxy-integrated flux density to be 85.5 mJy. As a part of our MLVL survey, we observed NGC3344 at 1.3 GHz using uGMRT and estimated galaxy-integrated flux density to be 7$\pm$0.05 mJy (\ref{gwb_contours_3344}). 
 Such a low value of flux density is likely due to missing short spacing needed to image the object at 1.3 GHz with uGMRT, which extends to almost 10'.
The total flux density of NGC3344 from our 0.4 GHz observation is 265$\pm$14 mJy. The low-resolution image at 0.4 GHz (Figure \ref{gwb_contours_3344}) detected far more diffuse emission beyond the disk of this galaxy than the high-resolution image which results in a significant difference in total fluxes from both the images.
As the 1.3 GHz uGMRT image could detect the central bright region, we could only estimate spectral indices for that central region (Figure \ref{gwb_contours_3344}). 

\begin{figure}
\centering
\includegraphics[trim={0 0 0 0},clip,scale=0.3,width=0.45\linewidth]{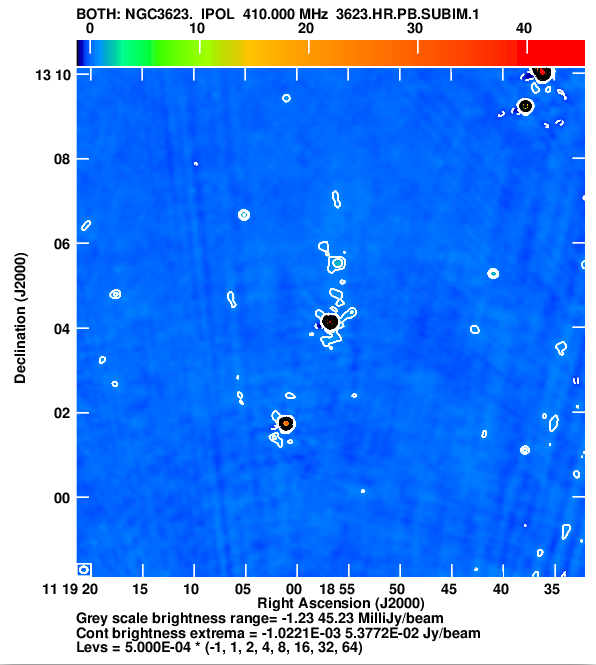}
\includegraphics[trim={0 0 0 0},clip,scale=0.3,width=0.45\linewidth]{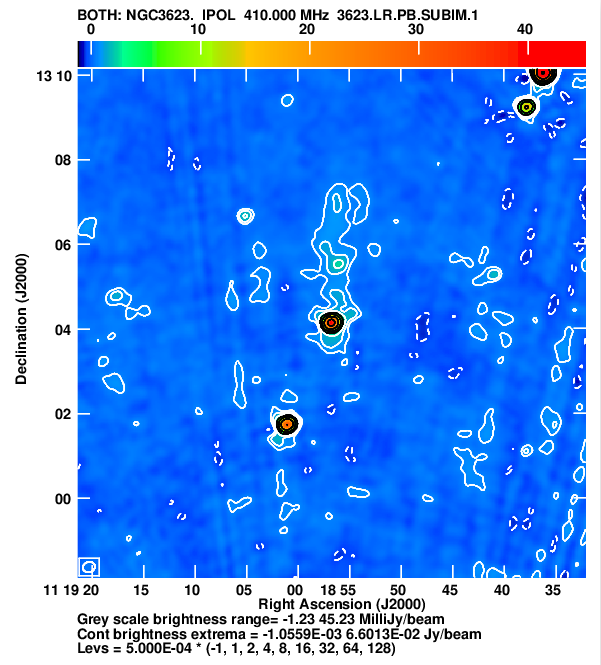}
\includegraphics[trim={0 0 0 0},clip,scale=0.3,width=0.45\linewidth]{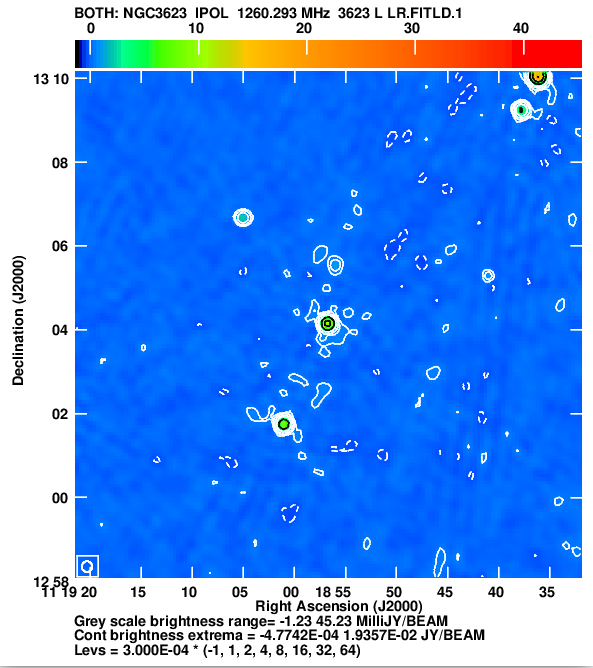}
\includegraphics[trim={0 0 0 0},clip,scale=0.3,width=0.45\linewidth]{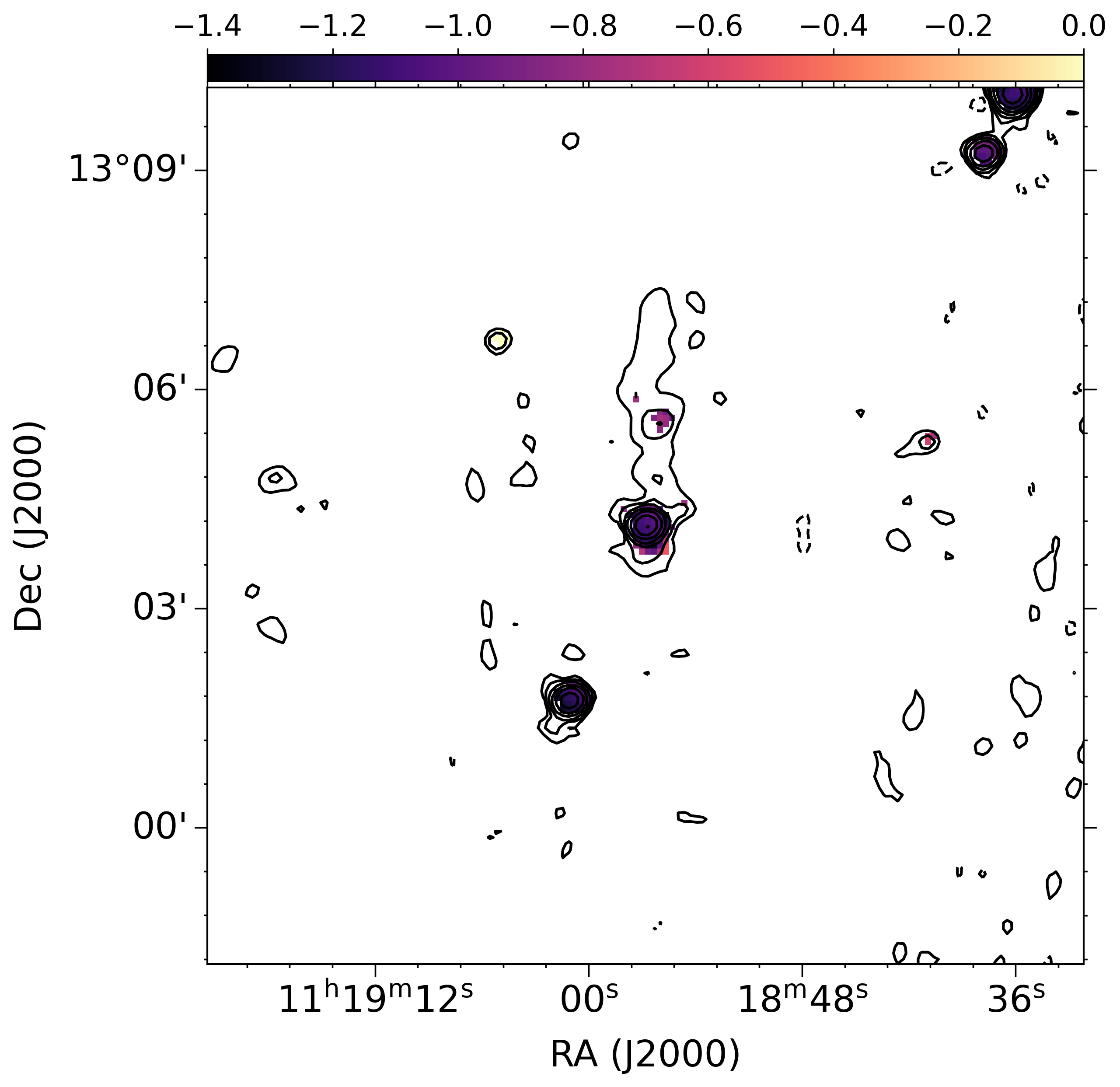} 
\caption{Top two panels show the high- and low-resolution image of NGC3623 at 0.4 GHz in contours and color scales. The contour levels are indicated below the figure. The resolution of the high- and low-resolution maps are provided in Table \ref{uGMRT_map_details}. The bottom left panel shows the uGMRT contour image at 1.3 GHz where the flux densities are shown in colors and contours. The bottom right panel shows the spectral index map between 0.4 GHz and 1.3 GHz in colors overlaid on the contours of the 0.4 GHz uGMRT image. The contour levels of the bottom right panel are plotted at 6.0$\times$10$^{-4}$ (-1, 1, 2, 4, 8, 16, 32, 64, 128, 256) Jy/beam. The spectral index map has a resolution of 17$\times$15 arcsec$^{2}$ and a mean error of 0.12.}
\label{gwb_contours_3623}
 \end{figure}

{\bf NGC3623} is a SAB(rs)a type of galaxy that is nearly edge-on with an inclination angle of 74 degrees \citep{Hogg2001AJ}. NGC3623 is interacting with two companion galaxies NGC3627 and NGC3628; these three galaxies are called Leo triplets \citep{Dumas2007MNRAS}. The galaxy-integrated flux density was estimated to be 9.2 mJy at 1.49 GHz using the VLA \citep{Condon1987}. 
We observed this galaxy at 1.3 GHz using uGMRT (Figure \ref{gwb_contours_3623}) and estimated a galaxy-integrated flux density of 20$\pm$8 mJy.
An elongated edge-on disk, of size $\approx$4 arcmin in the north-south direction, has been detected from our uGMRT observation with a total flux density of 78$\pm$20 mJy. We used the 1.3 and 0.4 GHz uGMRT maps to estimate the spectral index map of NGC3623. We could only estimate spectral index values of the compact regions due to the lack of extended emission in the 1.3 GHz image. 

\begin{figure}
\centering
\includegraphics[trim={0 0 0 0},clip,scale=0.3,width=0.45\linewidth]{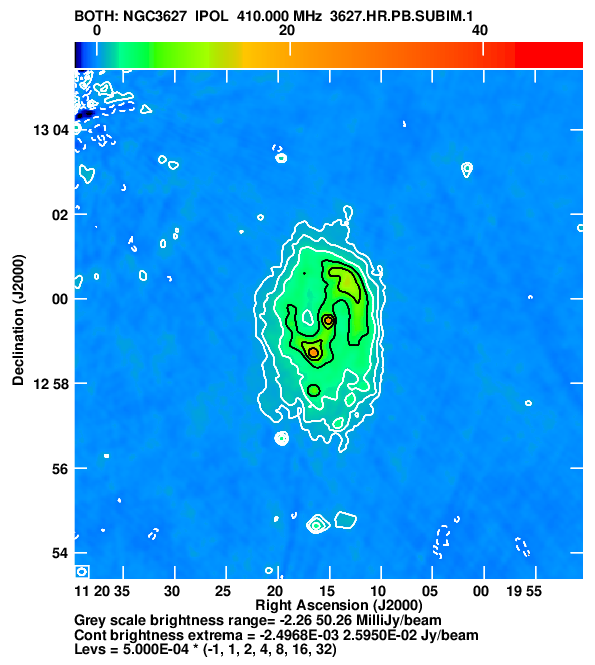}
\includegraphics[trim={0 0 0 0},clip,scale=0.3,width=0.45\linewidth]{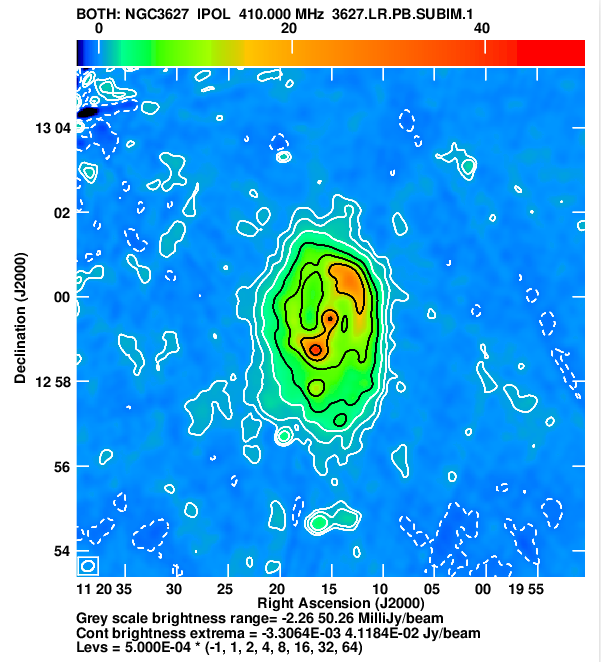}
\includegraphics[trim={0 0 0 0},clip,scale=0.3,width=0.45\linewidth]{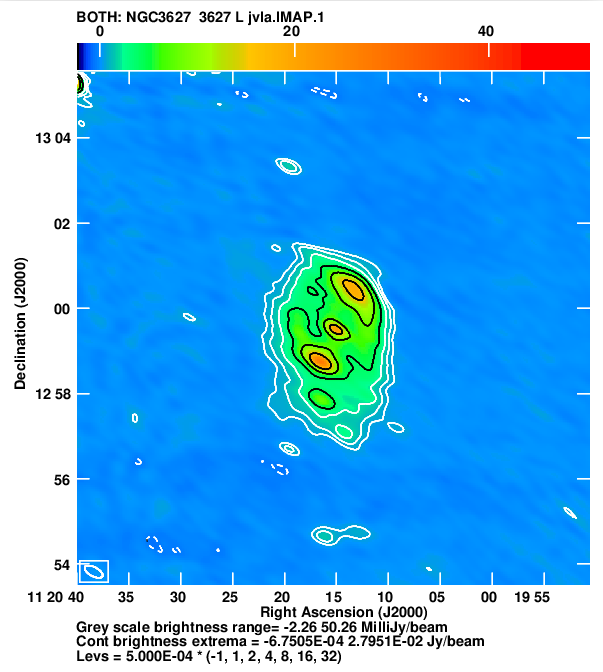}
 \includegraphics[trim={0 0 0 0},clip,scale=0.3,width=0.45\linewidth]{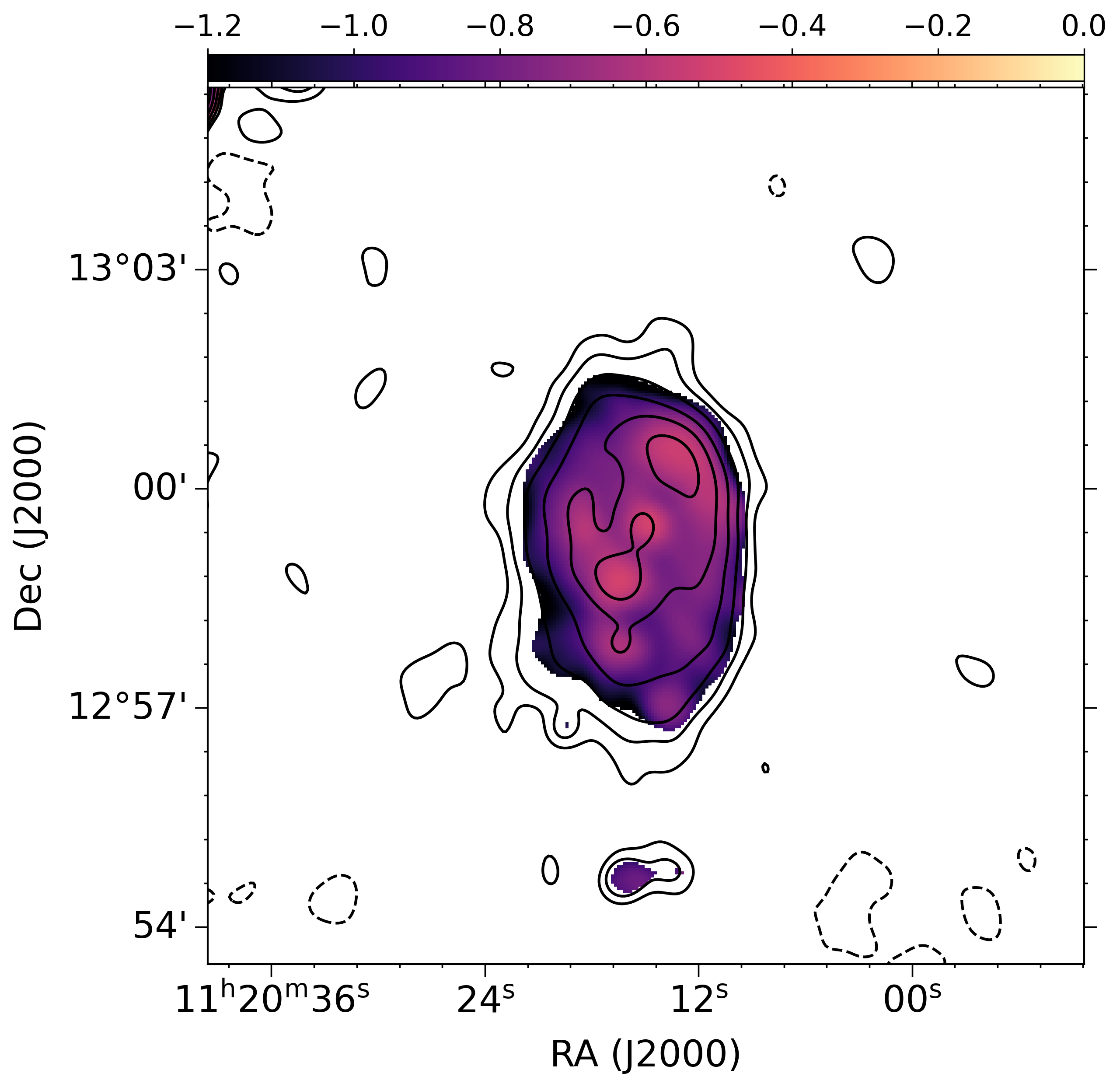} 
\caption{High- and low-resolution image of NGC3627 at 0.4 GHz in contours and color scales. The contour levels are indicated below the figure. The resolution of the high- and low-resolution maps are provided in Table \ref{uGMRT_map_details}. The bottom left panel shows the VLA contour image at 1.5 GHz where the flux densities are shown in colors and contours. The bottom right panel shows the spectral index map between 0.4 GHz and 1.4 GHz in colors overlaid on the contours of the 0.4 GHz uGMRT image. The contour levels of the bottom right panel are plotted at 1.5$\times$10$^{-3}$ (-1, 1, 2, 4, 8, 16, 32, 64, 128, 256) Jy/beam. The spectral index map has a resolution of 28$\times$28 arcsec$^{2}$ and a mean error of 0.05.}
\label{gwb_contours_3627}
\end{figure}

{\bf NGC3627} is another galaxy (type SAB) in the Leo group \citep{Dumas2007MNRAS}, which is inclined with an angle of 65 degrees with the sky plane. \cite{Paladino2009A&A} observed this galaxy at 0.327 MHz using the VLA. The sensitivity of the observation was very poor with an rms noise of 2 mJy/beam. We analyzed archival VLA wideband data at 1.5 GHz and found rms noise and total flux density of 150 $\mu$Jy/beam and 400$\pm$40 mJy respectively. A recent study of \cite{RoyManna2021} achieved an rms noise of 0.8 mJy/beam using legacy GMRT 0.33 GHz receivers. The authors found a galaxy-integrated flux density of 1.7$\pm$0.1 Jy and detected a diffuse halo of size $\approx$6 arcmin. We achieved significantly better rms noise (120 $\mu$Jy/beam) from our uGMRT observation at 0.4 GHz compared to the legacy GMRT observation. We estimated a galaxy-integrated flux density of 1.37$\pm$0.14 Jy.
Spectral index values steepen up to a value of -1 at the outer region of the galaxy (see Figure \ref{gwb_contours_3627}).

\begin{figure}
 \centering
\includegraphics[trim={0 0 0 0},clip,scale=0.3,width=0.45\linewidth]{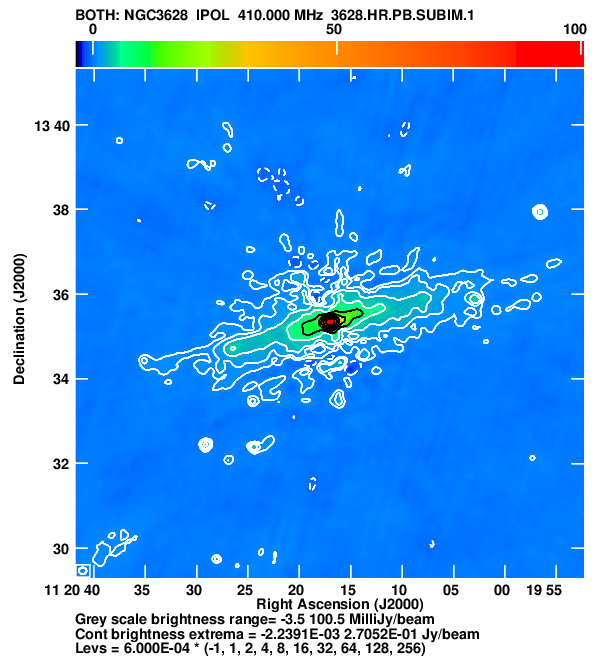}
\includegraphics[trim={0 0 0 0},clip,scale=0.3,width=0.45\linewidth]{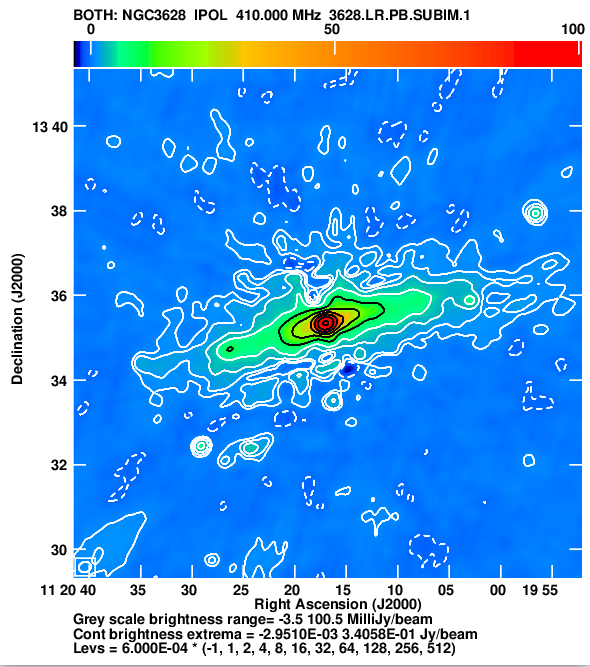}
\includegraphics[trim={0 0 0 0},clip,scale=0.3,width=0.45\linewidth]{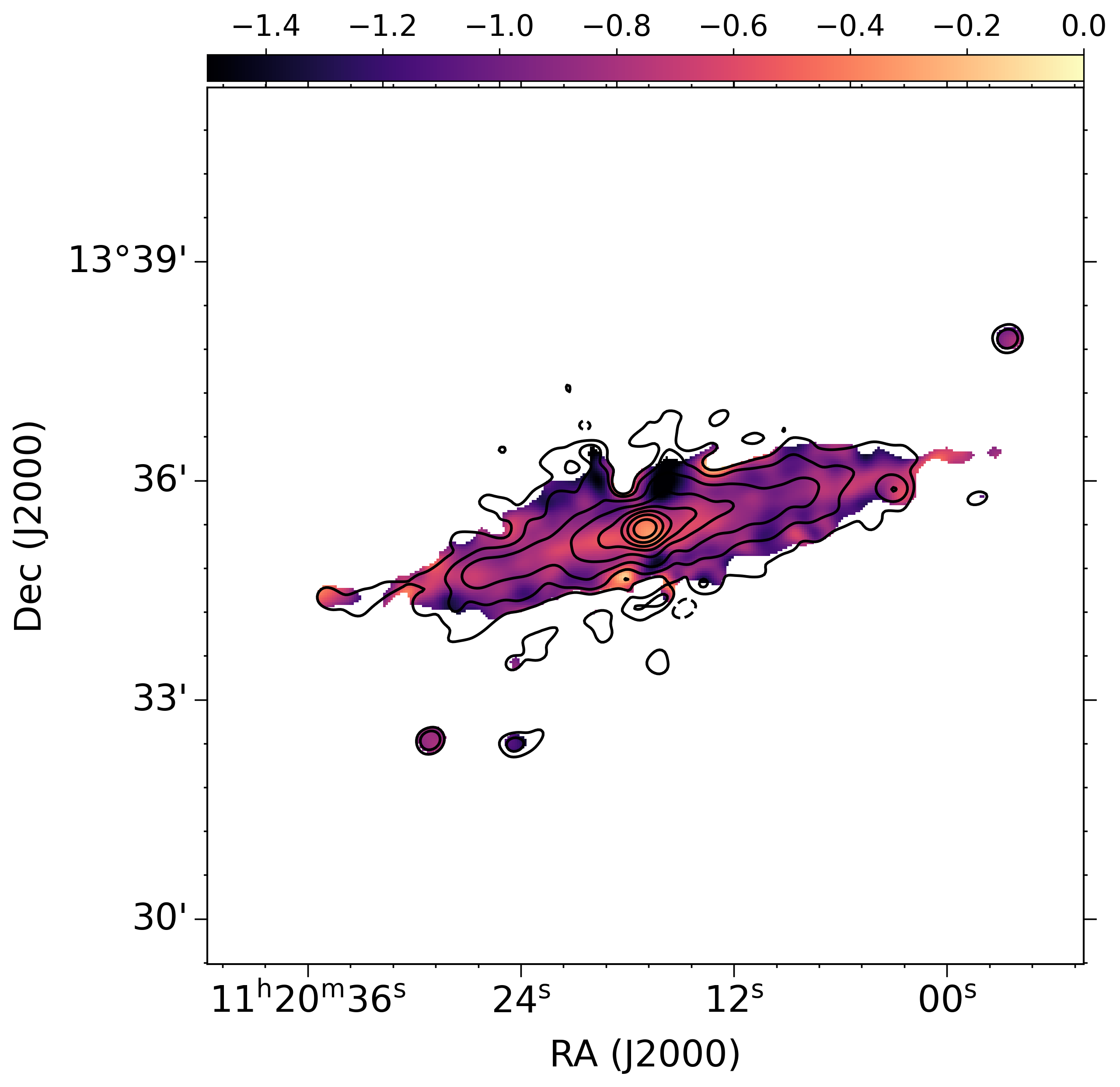}
\caption{Top two panels show the high- and low-resolution image of NGC3628 at 0.4 GHz in contours and color scales. The contour levels are indicated below the figure. The resolution of the high- and low-resolution maps are provided in Table \ref{uGMRT_map_details}. The bottom panel shows the spectral index map between 0.4 GHz and 1.5 GHz \citep{Irwin2012AJ144} in colors overlaid on the contours of the 0.4 GHz uGMRT image. The contour levels of the bottom panel are plotted at 1.5$\times$10$^{-3}$ (-1, 1, 2, 4, 8, 16, 32, 64, 128, 256) Jy/beam. The spectral index map has a resolution of 17$\times$15 arcsec$^{2}$ and a mean error of 0.1.}
\label{gwb_contours_3628}
\end{figure}

{\bf NGC3628} is the third member of the Leo triplet. NGC3628 is an Sb-type starburst galaxy inclined at an angle of $\approx$88 degree (edge-on) \citep{Baes2016A&A}. A large bridge (or tidal tail) of neutral gas extending toward the east from this galaxy has been observed which is believed to carry material from the galaxy to the intergalactic region \citep[e.g.][]{Hughes1991ApJ}. 
Radio observation of NGC3628 was carried out using VLA 1.49 GHz receivers \citep{Condon1987} and a galaxy-integrated flux density was measured to be 525 mJy. 
This galaxy was later observed using the Westerbork Synthesis Radio Telescope (WSRT) to find a total flux density of 590$\pm$10 mJy at $\approx$1.3 GHz \citep{Braun2007A&A}.
\cite{Nikiel2013A&A} observed the Leo triplet at 2.64 GHz using the 100-m Eﬀelsberg radio telescope to find a galaxy-integrated flux density of 364$\pm$17 mJy. The authors could not detect faint diffuse emissions from the galaxy due to the poor (1.0 mJy/beam) sensitivity of the observation.
NGC3628 was observed in the high-sensitive CHANG-ES survey \citep{Wiegert2015AJ} and the authors reported the total flux density to be 527.5$\pm$10.5 and 184.6$\pm$3.7 mJy at 1.5 and 6 GHz, respectively. We achieved an rms noise of $\sim$160 $\mu$Jy/beam in the surrounding region of NGC3628. We measure a total flux density of 1.2$\pm$0.25 Jy from the 0.4 GHz uGMRT map. We also estimate a mean spectral index of -0.86 (Figure \ref{gwb_contours_3628}) using the 1.5 GHz VLA map in C-array configuration and the 0.4 GHz uGMRT map respectively.


{\bf NGC4096} is a nearly edge-on galaxy (inclination angle: 76 degrees) which has a morphological type of SABc. \cite{Condon1987} estimated a galaxy-integrated flux density of 52.2 mJy at 20 cm. \cite{Wiegert2015AJ} observed this galaxy using VLA and measured its flux density to be 57.1$\pm$1.1 and 16.3$\pm$0.3 mJy at 19 and 5 cm, respectively. Recently, \cite{RoyManna2021} used the legacy GMRT and estimated a total flux density of 174$\pm$10 mJy at 0.33 GHz. The authors also measured the galaxy-integrated spectral index of -0.78$\pm$0.06 using 1.4 and 0.33 GHz observations. Our uGMRT observation of NGC4096 provides us with a total flux density of 137$\pm$12 mJy. Similar to NGC3627, the frequency difference of the legacy GMRT and uGMRT observation is the likely cause for the difference in the galaxy-integrated flux densities. We estimate the galaxy-integrated spectral index of -0.76$\pm$0.03  from 1.5 and 0.4 GHz observations and -0.65$\pm$0.06 from 1.5 and 0.144 GHz observations. 

{\bf NGC4594} is a SA(a) type spiral galaxy, also called a Sombrero galaxy, which has a large bulge and a prominent stellar disk \citep{Kormendy2004ARA&A}. The inclination angle of this galaxy is close to 90 degrees (edge-on) \citep{Jardel2011ApJ} and therefore, is an excellent candidate for studying disk-halo properties. NGC4594 was also observed to consist of (1) a massive black hole (mass $\sim$5.8$\times$ 10$^{8}$) \citep{Kormendy1996Apj}, and (2) an Active Galactic Nuclei (AGN) \citep{Gallimore2006AJ}. \cite{Kharb2016MNRAS} used the legacy GMRT 325 and 610 MHz receivers to study the kilo-parsec scale faint structure of NGC4594 due to the AGN activity. The authors achieved a poorer rms noise ($\sim$150$\mu$Jy/beam at 0.325 GHz) of the field than our observation ($\sim$23$\mu$Jy/beam at 0.4 GHz) and the faint diffuse emission was therefore not detected. Our deep uGMRT observation reveals a significant diffuse emission of NGC4594 which was not seen before. We estimated a total flux density of 118$\pm$37 mJy from the 0.4 GHz map. The galaxy-integrated spectral index is found to be -0.78$\pm$0.1 from 1.5 and 0.4 GHz observations. 

{\bf NGC4631} is a nearly edge-on galaxy with an inclination angle of 86 degrees. This galaxy is believed to interact with two companion galaxies, (1) a dwarf elliptical galaxy NGC4627 (towards the north-west) and, (2) a large edge-on galaxy NGC4656 (towards the south-east) \citep{Irwin2011MNRAS}. Previous studies also found a few more companion galaxies (e.g. NGC4631 Dw A) which are very faint \citep{Rand1994A&A, Rand1996AJ, Seth2005AJ}. All these interactions result in (1) a thick stellar disk and, (2) a multi-phase halo for the galaxy NGC4631 \citep{Seth2005AJb}.

\begin{figure}
 \centering
\includegraphics[trim={0 0 0 0},clip,scale=0.3,width=0.43\linewidth]{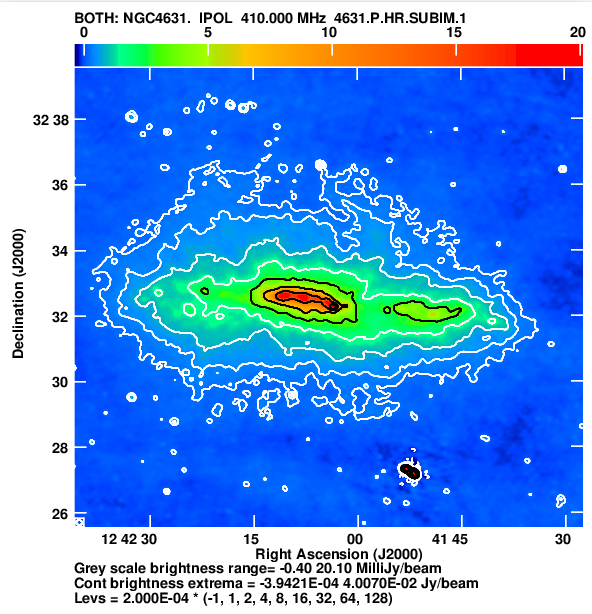}
\includegraphics[trim={0 0 0 0},clip,scale=0.3,width=0.4\linewidth]{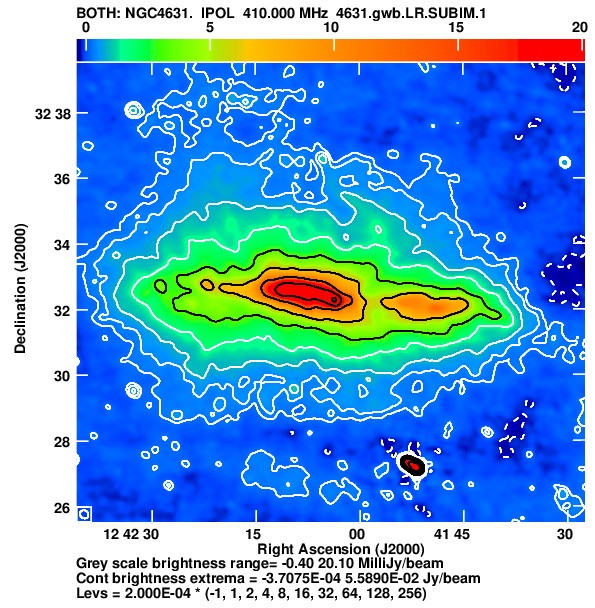}
\includegraphics[trim={0 0 0 0},clip,scale=0.3,width=0.4\linewidth]{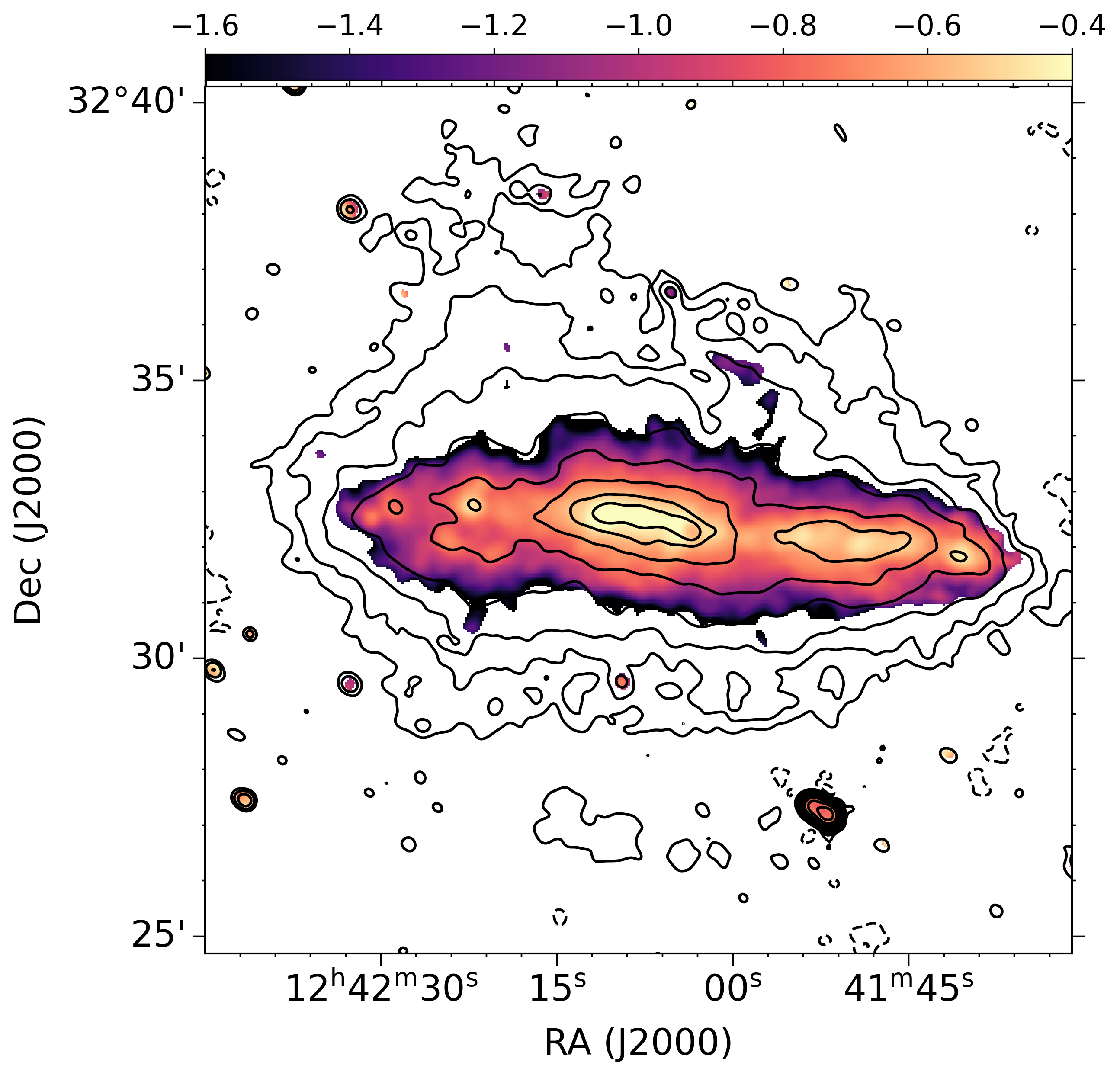}
\includegraphics[trim={0 0 0 0},clip,scale=0.3,width=0.4\linewidth]{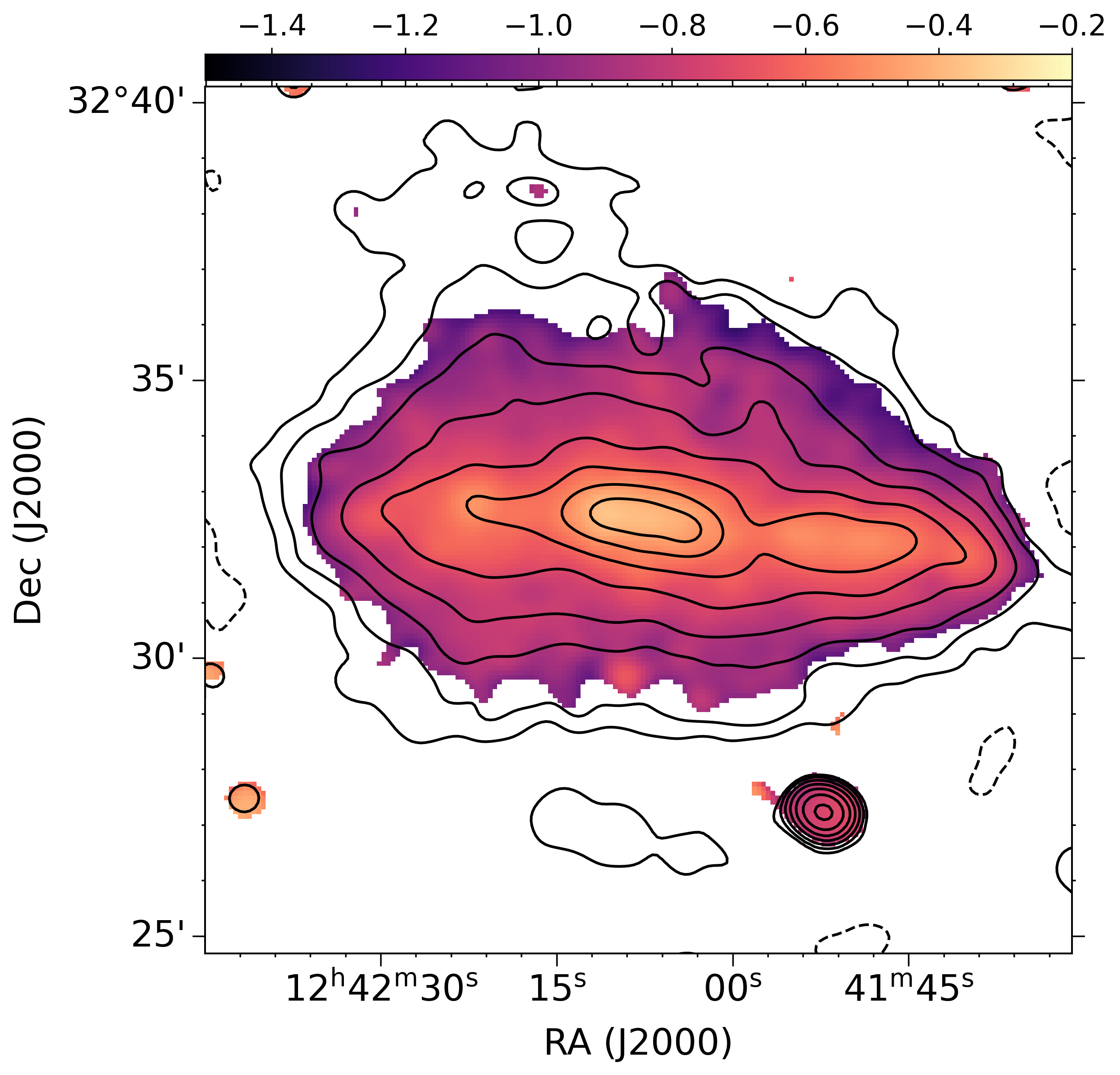}

\caption{Top two panels show the high- and low-resolution image of NGC4631 at 0.4 GHz in contours and color scales. The contour levels are indicated below the figure. The resolution of the high- and low-resolution maps are provided in Table \ref{uGMRT_map_details}. The bottom left panel shows the spectral index map using 0.4 GHz uGMRT and 1.5 GHz VLA image \citep{Irwin2012AJ144}, in colors overlaid on the contours of the 0.4 GHz uGMRT image. The contour levels of the bottom left panel are plotted at 2.25$\times$10$^{-4}$ (-1, 1, 2, 4, 8, 16, 32, 64, 128, 256) Jy/beam. The bottom right panel shows the spectral index map using 0.144 GHz \citep{Shimwell2022A&A} and 1.5 GHz maps, in colors overlaid on the contours of the 0.4 GHz uGMRT image. The contour levels of the bottom right panel are plotted at 8.0$\times$10$^{-4}$ (-1, 1, 2, 4, 8, 16, 32, 64, 128, 256) Jy/beam. The spectral index maps have resolutions of 18$\times$15 arcsec$^{2}$ and 34$\times$33 arcsec$^{2}$ respectively. The spectral index maps have mean errors 0.03 and 0.06 respectively.}
\label{gwb_contours_4631}
  \end{figure}

VLA observations at 1.5 and 6 GHz in its C configuration \citep{Irwin2012AJ} detected a large halo of this galaxy of size $\sim$5 arcmin. The authors estimated a total flux density of 935 mJy at 1.5 GHz and an average spectral index of -0.84$\pm$0.05. This galaxy has also been observed by LOFAR at 144 MHz by \cite{Shimwell2022A&A} and a total flux density of 5.9 Jy has been estimated. The authors achieved an rms noise of $\approx$160$\mu$Jy/beam and a resolution of 20arcsec$\times$20arcsec. 
We have achieved an rms noise of $\approx$50 $\mu$Jy/beam
from our high-resolution map at 0.4 GHz. The measured flux density from our 0.4 GHz map is 2.69$\pm$0.14 Jy. The spatially resolved spectral index maps (Figure \ref{gwb_contours_4631}) show a steepening of the spectral index value up to -1.5 at the outer region of the galaxy with a galaxy-integrated spectral index value of $\sim$-0.9. Note that the galaxy-integrated spectral index was found to be -0.73 and -0.84 
by \cite{Hummel1990A&A} and \cite{Irwin2012AJ} respectively. 

\section{Discussion}
\label{discussion}
Magnetic fields and CREs are key components to probe various physical processes in the ISM such as outflows, disk-halo connections, X-shaped structure of the magnetic ﬁelds, and propagation of CREs. Understanding these processes from local to galaxy-integrated scales is crucial to know the evolution of different types of galaxies. Most of the existing radio observations are not sensitive enough to address the issues mentioned above. However, VLA and LOFAR recently provided us with deep observations of some nearby galaxies as part of different surveys such as CHANG-ES and LoTSS respectively. 
In this section, we discuss the findings of the whole sample and also from each individual galaxy to better understand if there exists any association of any physical process with the galaxy type.
In Section \ref{Radio_Continuum_Halos}, we present our study on the detected radio halos of 5 nearly edge-on galaxies. In Section \ref{Diffuse_Emission}, we discuss the study on the 2 Galaxies with moderate inclination angles.
We discuss the magnetic fields and the propagation of CREs of the sample galaxies in Section \ref{discussion_CRE_B}.

\subsection{Radio Continuum Halos of 5 Edge-on Galaxies}
\label{Radio_Continuum_Halos}

We aimed to compare the distributions of the radio emissions of our low-resolution uGMRT maps at 0.4~GHz with the VLA or uGMRT maps at $\sim$1.5~GHz. To get the same resolution and pixel size for both the maps, we convolved and re-grided the uGMRT maps to the same beam sizes and pixel sizes as the maps at $\sim$1.5~GHz. Thereafter, we used \textit{CASA viewer} to extract the flux densities along two straight line directions from the galaxy centre; (1) parallel to the galaxy disk and (2) perpendicular to the galaxy disk. We took these $``$cross-cuts$"$ along the same straight lines for both maps at 0.4 and $\sim$1.5~GHz.

We used the cross-cuts to characterize the size of the radio continuum halos along the disk and perpendicular to the disk. We measured the distance from the galaxy centre to the location where the flux density value drops to 5 times the rms noise in the surrounding region of a galaxy. It provides a good estimate of the galaxy sizes, but the sizes could depend on the sensitivity of the measurements. In Table \ref{halo_sizes}, we present the measured sizes of the radio continuum halos, at 0.4 and $\sim$1.5 GHz frequencies, estimated using the cross-cuts in both directions. 
For the first time, we detected a disk of $\sim$8 kpc of NGC3623 at 0.4 GHz. The vertical halo of NGC3623 was also detected to be $\approx$30\% larger at 0.4 GHz than compared to the halo at 1.3 GHz. 
For the galaxy NGC4096, the halo sizes in the uGMRT image are more than 70\% and 50\% larger than the sizes in the VLA map at 1.5 GHz image along the disk and perpendicular to the disk respectively. For the edge-on galaxy NGC4594, we could detect, for the first time, a radio disk of size $\approx$10 kpc whereas the VLA observation could detect only the central AGN. The halo size in the vertical direction is also $\sim$30\% larger at 0.4~GHz than the same at 1.5 GHz. Similarly, the halo sizes of NGC4631 at 0.4 GHz are $\approx$10\% and $\approx$75\% larger than the halos at 1.5 GHz along the disk and perpendicular to the disk respectively. We could not detect significant differences in the size of the radio halos of NGC3628, in both the perpendicular direction, at 0.4 GHz compared to the VLA image at 1.5 GHz.

We also compared the distribution of the radio halos at 0.4 and $\sim$1.5 GHz images by fitting an exponential function to the cross-cuts along the direction of the disk as well as perpendicular to the disk. We used the following form of the exponential function, an approximation to the function that describes how the intensity of a spiral disk galaxy varies with distance from its centre, to get the position where the flux density drops to $\approx$36 percent of the peak value. 

\begin{equation}
    f(x)=a \times exp(-x/b)
\end{equation}

In Table \ref{fitted_para_sample}, we present the b values along the disk and perpendicular to the disk for these 5 edge-on galaxies at 0.4 and $\sim$1.5 GHz. For the 4 galaxies except NGC4631, the distributions of radio halos along the disks are significantly (with $>$6 sigma significance) larger at 0.4 GHz than the radio halo at 1.5 GHz. For NGC4631, the b value along the disk is higher than the same at 1.5 GHz with 2 sigma significance.  
Note that we could detect elongated disks at 0.4 GHz for NGC3623 and NGC4594; therefore, the b values are much higher at 0.4 GHz than the 1.5 GHz image. In the direction perpendicular to the disk, the b values at 0.4 GHz are higher than 1.5 GHz with statistical significance of 8.1 and 7.6 sigma for NGC4096 and NGC4631 respectively. For NGC3623, NGC3628, and NGC4594, b values are similar at both frequencies.

\begin{table}
 \caption{Sizes of radio halos of the VLA and the uGMRT images. These sizes are estimated from the galaxy's centre to the point where flux density drops to 5 times the mean rms noise in the surrounding region of the concerned galaxy.}
\scriptsize
\centering
 \begin{tabular}{||c c c c c||} 
 \hline
Name & Along the disk & Along the disk & Perpendicular to the disk & Perpendicular to the disk \\ 
\hline
     &  VLA/uGMRT image &  uGMRT image  &  VLA/uGMRT image &  uGMRT image \\   
     & at $\sim$1.5 GHz  & at 0.4 GHz    & at $\sim$1.5 GHz & at 0.4 GHz \\
     & (kpc)  & (kpc)    & (kpc) & (kpc) \\
     
 \hline
NGC3623      & 0.65 & 4.33 & 0.64 & 0.85 \\
NGC3628     & 11.00  & 13.20 & 2.58 & 2.67   \\
NGC4096     & 4.10  & 6.27 & 1.45 &  2.50     \\
NGC4594     & 0.63  & 5.79 & 1.26 &  1.63     \\
NGC4631     &  12.99  & 14.48 & 3.99 & 6.97     \\
 \hline
 \end{tabular}
\label{halo_sizes}
\end{table}

     

\begin{table}
 \caption{The b values, where the flux density drops to $\approx$36 percent of the peak value (Equation 1), along the disk and perpendicular to the disk for the VLA and the uGMRT images.}
\scriptsize
\centering
 \begin{tabular}{||c c c c c||} 
 \hline
Name & Along the disk & Along the disk & Perpendicular to the disk & Perpendicular to the disk \\ 
\hline
     &  VLA/uGMRT image &  uGMRT image  &  VLA/uGMRT image &  uGMRT image \\   
     & at $\sim$1.5 GHz  & at 0.4 GHz    & at $\sim$1.5 GHz & at 0.4 GHz \\
     & (kpc)  & (kpc)    & (kpc) & (kpc) \\
     
 \hline
NGC3623     & 1.19$\pm$0.15 & 4.12$\pm$0.57 & 0.46$\pm$0.05 & 0.53$\pm$0.06     \\
NGC3628     &  0.48 $\pm$0.01  & 0.58$\pm$0.01 & 0.39$\pm$0.02 & 0.40$\pm$0.02   \\
NGC4096     &  1.55$\pm$0.06  & 2.24$\pm$0.08 & 0.55$\pm$0.02 &  0.78$\pm$0.02     \\
NGC4594     &  0.12$\pm$0.01  & 3.99$\pm$0.31 & 0.33$\pm$0.01 &  0.33$\pm$0.01     \\
NGC4631     &  3.97$\pm$0.31  & 4.87$\pm$0.30 & 0.57$\pm$0.01 &  0.74$\pm$0.02    \\
 \hline
 \end{tabular}
\label{fitted_para_sample}
\end{table}

\subsection{Diffuse Emission of 2 Galaxies with Moderate Inclination}
\label{Diffuse_Emission}
We compared the faint diffuse emission observed by uGMRT at 0.4 GHz with that of the $\sim$1.5 GHz emissions of NGC3344 and NGC3627, which have inclination angles $\approx$65 degrees. We followed the same method as described in Section \ref{Radio_Continuum_Halos} to quantify the sizes and distributions of the radio disks in the sky-plane. For NGC3344, we took a cross-cut along the north-east direction from the centre of the galaxy and found that the sizes of the radio disk (where flux density drops to 5$\sigma$; $\sigma$ is the rms noise) are 0.98 and 5.28 kpc at 1.3 and 0.4 GHz respectively. For NGC3627, we took the cross-cut perpendicular to the galaxy bar, and the sizes of the radio disk are measured to be 4.57 and 4.96 kpc at 1.5 and 0.4 GHz respectively. Therefore, we could detect, (1) a significantly large radio halo of NGC3344 at 0.4 GHz compared to 1.3 GHz and (2) a comparable halo for NGC3627 at 0.4 and 1.5 GHz.   

The b values, where the flux density drops to 36 percent of the peak value, are measured to be 0.69$\pm$0.06 and 2.38$\pm$0.15 kpc at 1.3 and 0.4 GHz for NGC3344 along the cross-cuts mentioned above. Similarly, for NGC3627, b values are estimated to be 2.22$\pm$0.13 and 2.90$\pm$0.20 kpc at 1.5 and 0.4 GHz respectively. Therefore, the distributions of the radio emission along these cross-cuts are found to be significantly smoother (significant by 10 and 3 sigma for NGC3344 and NGC3627 respectively) at 0.4 GHz compared to the emission at $\sim$1.5 GHz. 

\subsection{CRE Propagation and Magnetic Fields of the Sample Galaxies}
\label{discussion_CRE_B}

Observational evidence (e.g. extra-planer halo and X-shaped structure of magnetic fields) exist that support the export of materials (e.g. cold gas, CREs, etc.) from the galaxy disk to the outer region of the disk. Although the typical thickness of the galaxy disks of large spirals ranges between 0.3 to 0.5 kpc, recent radio observations with high sensitivity are providing vertical halo sizes that are much larger than 1 kpc \citep[e.g.][]{RoyManna2021, Irwin2012AJ, Wiegert2015AJ, Shimwell2022A&A}. In the plane of the disks, detected radio halos are also significantly larger than the size of the disk of some galaxies \citep[e.g.][]{RoyManna2021,Shimwell2017A&A}, providing evidence of escape of CREs from the host galaxy \citep[e.g. Non-calorimetric model,][]{Niklas1997A&A}. As presented in this paper, our uGMRT observations of these 7 MLVL galaxies provide unprecedented sensitivity at 0.4 GHz frequency which results in the detection of radio halos that are significantly larger than the galaxy disks and the wideband VLA observations (Section \ref{Radio_Continuum_Halos}). In Section \ref{Radio_Continuum_Halos}, we also demonstrated that the b values, where the flux density drops to 36\%, at 0.4 GHz are remarkably larger than the same at $\sim$1.5 GHz. A large b value at 0.4 GHz compared to the same at $\sim$1.5 GHz could occur if the mechanism of CRE propagation is dominated by the process of Streaming Instability \citep{RoyManna2021}, where CREs propagate along the magnetic field lines in regions of highly ordered magnetic fields. Note that the dominant mechanism of CRE propagation was found to be streaming instability for 3 out of 7 galaxies of our pilot study \citep{RoyManna2021}. Besides, CREs could be expelled outside the disk through the process outflows resulting in a large value of b. However, disentangling the dominant mechanism of CRE propagation would require detailed modelling of these processes using multi-frequency radio observations. In a future paper, we will present our findings on the dominant mechanism of CRE propagation of the MLVL galaxies.

As mentioned earlier, hints of extra-planer magnetic fields (e.g. X-shaped magnetic fields)
have been reported in the past and researchers are keen to investigate such components using deep observations at low radio frequencies. Such extra-planer components of magnetic fields are critical for understanding the evolution of galaxies. Synchrotron emission, an excellent tracer of magnetic field components, contributes $\sim$95\% to the total radio emission. Therefore, our uGMRT observations at 0.4 GHz essentially provide the signature of the magnetic fields of the sample galaxies. As shown in Figures \ref{gwb_contours_3344} - \ref{gwb_contours_4631} and presented in Section \ref{Radio_Continuum_Halos}, typical values of the vertical flux density at 0.4 GHz are found to be $\approx$1 mJy/beam at a distance of $\approx$0.8, $\approx$3.0, $\approx$1.4, $\approx$1.1 and $\approx$5.6 kpc (from the galaxy centre) for the 5 edge-on galaxy- NGC3623, NGC3628, NGC4096, NGC4594, and NGC4631 respectively. Considering typical values of the spectral index of -0.8, the above value of flux density ($\approx$1mJy) would correspond to an equipartition$\footnote{We can only estimate magnetic fields at the equipartition condition (or the minimum energy condition) when the total energy in magnetic fields and CREs is minimum.}$ magnetic field \citep{BeckandKrause2005} value of $\approx$20~$\mu$G. Value of the equipartition magnetic field would be even higher for a steeper spectral index value (e.g. $\approx$30 $\mu$G for a spectral index value of -1.5). 
Therefore, our uGMRT observations indicate a high value of equipartition magnetic fields ($\sim$20-30 $\mu$G) at vertical distances of more than a few kpcs from the centres of the sample galaxies. 
Few observational evidences exist \citep[e.g.][]{Irwin2024,Krause2020A&A,Shimwell2022A&A} that detected large radio halos of edge-on galaxies and predicted high values of magnetic fields at few kpc distances. \cite{Stein2020A&A} studied the galaxy NGC4217 and found large scale coherent magnetic fields at a distance of $\sim$7 kpc. However, magnetic field values are not known that far away from galactic disk.
Different models exist such as galactic chimneys \citep{Norman1989ApJ}, galactic fountains \citep{Shapiro1976ApJ}, super-bubble blowout \citep{MacLow1999ApJ}, CR-driven Parker instability \citep{Hanasz2002A&A} and galactic dynamo including outflow and accretion \citep{Woodfinden2019MNRAS} that could explain the above-mentioned observations including disk-halo connections, X-shaped structure of the magnetic fields, and larger propagation scales of CREs in the direction perpendicular to the disk.
Detailed modelling is required for a better understanding of such high values of magnetic fields in the halo regions of nearby galaxies. In a future paper, we will present our study on spatially-resolved magnetic fields of the MLVL galaxies to better understand the physical processes at sub-kpc linear scales.


\section{Summary}
\label{summary}
\begin{enumerate}
    \item We utilized the large bandwidth of uGMRT to observe 7 galaxies from our MLVL survey at 0.3-0.5 GHz. We made high- and low-resolution images of the 7 sample galaxies with sensitivities $\approx$3-4 times higher than the images made using the legacy GMRT. To our knowledge, these images that are the most sensitive ones at these frequencies. 
    For the first time, we detected elongated radio disks for the edge-on galaxy NGC3623 and NGC4594.

    \item The detected halo sizes in the vertical direction of 4 edge-on MLVL galaxies- NGC3623, NGC4096, NGC4594, and NGC4631 at 0.4 GHz are
     larger by 30\%, 50\%, 30\%, and 75\% respectively than their images at $\sim$1.5 GHz. In the direction of the disks, the halo sizes of these 4 uGMRT images are also significantly larger than images at 1.5 GHz. Such large halo sizes of these galaxies are being reported for the first time. For NGC3628, the halo sizes at 0.4 GHz are comparable to the same at 1.5 GHz in both the perpendicular directions.

    \item We also compared the distribution of radio emission of 0.4 GHz uGMRT images with the images at $\sim$1.5 GHz by fitting an exponential function to the flux densities along cross-cuts in the plane of the disk and perpendicular to the disk. The distributions are found to be remarkably larger at 0.4 GHz than compared to 1.5 GHz.

    \item For NGC3344, the size of the radio disks in the sky-plane is measured to be significantly larger at 0.4 GHz than compared to 1.3 GHz. For NGC3627, halo sizes are comparable at both frequencies. The radio emissions are also remarkably smooth at 0.4 GHz for both galaxies.

    \item We made spatially-resolved spectral index maps of all the sample galaxies using uGMRT or VLA images at 1.5 GHz and uGMRT images at 0.4 GHz. We also made spectral index maps of NGC4096 and NGC4631 using 0.4 GHz uGMRT maps and 0.144 GHz LOFAR maps observed as a part of the LoTSS survey. Our study demonstrates how spectral indices of these galaxies steepen up to a value of $\approx$-1.5 at the halo regions.
    
   \item We find evidence of high magnetic fields ($\sim$20 $\mu$G) typically found on galactic disks even $>$5 kpc away from galactic disk in one of the galaxies studied (NGC 4631). To our knowledge, such a high magnetic field strength has not been reported in literature that far away from galactic disk.

\end{enumerate}

\begin{acknowledgments}
We would like to thank Aditya Chowdhury for help in analyzing the uGMRT data.
We thank Judith Irwin who provided their earlier published images at 1.5 GHz. We also thank the anonymous referee whose comments helped signiﬁcantly improve the presentation of the paper.
We thank the staff members of the S. N. Bose National Centre for Basic Sciences under the Department of Science and Technology, Govt. of India, for their help in our work.
We thank the staff of GMRT that allowed these observations to be made. GMRT is run by National Centre for Radio Astrophysics of the Tata Institute of Fundamental Research. We acknowledge the support of the Department of Atomic Energy, Government of India, under project No. 12-R$\&$D-TFR-5.02-0700.
\end{acknowledgments}

\bibliography{main.bib}{}
\bibliographystyle{aasjournal}

\end{document}